\documentclass[12pt,draftclsnofoot,onecolumn]{IEEEtran}
\makeatletter
\newcommand\semihuge{\@setfontsize\semihuge{22.3}{22}}
\makeatother
\usepackage{algpseudocode}
\usepackage{algorithm}
\usepackage{algorithmicx}

\usepackage{lipsum} % For algorithm smape after or before:
\usepackage{amsfonts}

\usepackage[dvips]{color}
\usepackage{comment}
\usepackage{todonotes}
\usepackage{epsf}
\usepackage{epsfig}
\usepackage{times}
\usepackage{epsfig}
\usepackage{graphicx}
\usepackage{mathtools}
\usepackage{mathrsfs}
\usepackage{amssymb}
\usepackage{pdfpages}
\usepackage{epstopdf}
\usepackage{float}
\newfloat{algorithm}{t}{lop}
\usepackage{subfig}
\usepackage{dsfont}
\usepackage[multiple]{footmisc}
\usepackage{lettrine} % \lettrine[findent=1pt]{{{R}}}{}
\usepackage{amsmath,epsfig,amssymb,algorithm,algpseudocode,amsthm,cite,url}
\usepackage{caption}
\allowdisplaybreaks
\usepackage{csquotes}
\newcommand{\R}{\ensuremath{{\mathbb R}}}
\newcommand{\E}{\ensuremath{{\mathbb E}}}

\usepackage{multirow}
\DeclareMathOperator*{\argmax}{arg\,max}

\usepackage{verbatim}
\usepackage{subfig}
\usepackage[english]{babel}
\usepackage{amsmath,amssymb}

\usepackage{verbatim}

\newtheorem{corollary}{Corollary}
\newtheorem{theorem}{\bf Theorem}

\newtheorem{proposition}{\bf Proposition}
\newtheorem{lemma}{\bf Lemma}

\begin{document}
\title{\semihuge  Joint Communication and Control for Wireless Autonomous Vehicular Platoon Systems \vspace{-0.5cm}}

\author{\IEEEauthorblockN{ Tengchan Zeng$^1$, Omid Semiari$^2$, Walid Saad$^1$, and Mehdi Bennis$^3$}\vspace{-0.05cm}\\
	\IEEEauthorblockA{
		\small $^1$ Wireless@VT, Department of Electrical and Computer Engineering, Virginia Tech, Blacksburg, VA, USA,\vspace{-0.1cm}\\ Emails:\url{{tengchan , walids}@vt.edu}.\vspace{-0.1cm}\\
		$^2$ Department of Electrical and Computer Engineering, University of Colorado Colorado Springs, Colorado Springs, CO, USA,\vspace{-0.1cm} \\ Email: \url{osemiari@uccs.edu}.\vspace{-0.1cm}\\
		$^3$ Centre for Wireless Communications, University of Oulu, Oulu, Finland, Email:
		\url{mehdi.bennis@oulu.fi}.
		\thanks{\textcolor{black}{
				A preliminary version of this work appears in the proceeding of IEEE ICC, 2018 \cite{zeng2018integrated}. This research was supported by the U.S. National Science Foundation under Grants CNS-1513697, CNS-1739642, CNS-1941348, and IIS-1633363, as well as by the Academy of Finland project (CARMA), INFOTECH project (NOOR), and Kvantum Institute strategic project (SAFARI).}}\vspace{-0.90in}
		%\thanks{This work was supported by the U.S. National Science
		%Foundation under Grants AST-1506297, by the Office of Naval Research (ONR) under Grant N00014-15-1-2709, and, by the ERC Starting
		%Grant 305123 MORE (Advanced Mathematical Tools for Complex Network
		%Engineering), and by the Academy of Finland.}
}}
\maketitle

\begin{abstract}
	\vspace{-0.15in}
	Autonomous vehicular platoons will play an important role in improving on-road safety in tomorrow's smart cities.
	Vehicles in an autonomous platoon can exploit vehicle-to-vehicle (V2V) communications to collect environmental information so as to maintain the target velocity and inter-vehicle distance.
	However, due to the uncertainty of the wireless channel, V2V communications within a platoon will experience a wireless system delay.
	Such system delay can impair the vehicles' ability to stabilize their velocity and distances within their platoon.
	In this paper, the problem of integrated communication and control system is studied for wireless connected autonomous vehicular platoons.
	In particular, a novel framework is proposed for optimizing a platoon's operation while jointly taking into account the delay of the wireless V2V network and the stability of the vehicle's control system.
	First, stability analysis for the control system is performed and the maximum wireless system delay requirements which can prevent the instability of the control system are derived.
	Then, delay analysis is conducted to determine the end-to-end delay, including queuing, processing, and transmission delay for the V2V link in the wireless network.
	Subsequently, using the derived wireless delay, a lower bound and an approximated expression of the reliability for the wireless system, defined as the probability that the wireless system meets the control system's delay needs, are derived.
	Then, the parameters of the control system are optimized in a way to maximize the derived wireless system reliability.
	Simulation results corroborate the analytical derivations and study the impact of parameters, such as the packet size and the platoon size, on the reliability performance of the vehicular platoon. 
	More importantly, the simulation results shed light on the benefits of integrating control system and wireless network design while providing guidelines for designing an autonomous platoon so as to realize the required wireless network reliability and control system stability.

	%Simulation results corroborate the stability analysis and shed light on the benefits of jointly considering the design of the control system and the wireless network for {\color{blue}wireless} connected platoons.	
	%More importantly, the simulation results show that the proposed framework can provide guidelines on how to choose the number of follower, the spacing, the velocity, and the control parameters for the platoon 
	
\end{abstract} \vspace{0.1cm}

\section{Introduction}%\vspace{-0.01cm}
\vspace{-0.05in}
Intelligent transportation systems (ITSs) will be a major component of tomorrow's smart cities.
In essence, ITSs will provide a much safer and more coordinated traffic network
by using efficient traffic management approaches \cite{ferdowsi2017deep}. 
One promising ITS service is the so-called \emph{autonomous vehicular platoon system}, which is essentially a group of vehicles that operate together and continuously coordinate their speed and distance.
By allowing autonomous vehicles to self-organize into a platoon, the road capacity can increase so as to prevent traffic jams \cite{hall2005vehicle}. 
Also, vehicles in the platoon can raise the fuel efficiency \cite{liang2016heavy}.
Furthermore, platoons can provide passengers with more comfortable
trips, especially during long travels \cite{nowakowski2010cooperative}.

To reap the benefits of platooning, one must ensure that each vehicle in the platoon has enough awareness of its relative distance and velocity with its surrounding vehicles. 
This is needed to enable vehicles in a platoon to coordinate their acceleration and deceleration. 
In particular, enabling autonomous platooning requires two technologies:  vehicle-to-vehicle (V2V) communications \cite{perfecto2017millimeter} and cooperative adaptive cruise control (CACC) \cite{jin2013stability}. 
V2V communications enable vehicles to exchange information, such as high definition (HD) maps, velocity, and acceleration \cite{bergenhem2012overview}.
Meanwhile, CACC is primarily a control system that allows control of the distances between vehicles using information collected by sensors and V2V links. Effectively integrating the operation of the CACC system and the V2V communication network is central for successful platooning in ITSs. 
%Based on the collected information, the vehicles can utilize control mechanism to determine whether to accelerate or brake and how much force the engine needs to provide. 

Nevertheless, due to the uncertainty of the wireless channel, V2V communication links will inevitably suffer from time-varying delays, which can be as high as hundreds of millisecond \cite{xu2017dsrc}.
Unfortunately, if the delayed information is used in the design of the autonomous vehicles' control system, such information can jeopardize the stable operation of the platoon \cite{liu2001effects}. 
Therefore, to maintain the stability of a platoon, the control system must be robust to such wireless transmission delays. 
To this end, one must jointly consider the control and wireless systems of a platoon to guarantee low latency and stability. 

The prior art on vehicular platooning can be grouped into two categories.
The first category focuses on the performance analysis, such as interference management \cite{peng2017resource,bithas2017double,bithas2017transmit,du2017stacked}, security system design \cite{xu2017surveillance}, and transmission delay analysis  \cite{peng2017performance,hu2017end,kaewpuang2017cooperative}, for the inter-vehicle communication network. 
The second category designs control strategies that guarantee the stability of the platoon system. 
Such strategies include adaptive cruise control (ACC) \cite{vahidi2003research}, enhanced ACC \cite{li2004modeling}, and connected cruise control (CCC) \cite{orosz2016connected}. 
However, these works are limited in two aspects. The communication-centric works in \cite{peng2017resource,bithas2017double,bithas2017transmit,du2017stacked,xu2017surveillance,peng2017performance,hu2017end,kaewpuang2017cooperative} completely abstract the control system and do not study the impact of wireless communications on the platoon's stability. 
Meanwhile, the control-centric works in \cite{li2004modeling,vahidi2003research,orosz2016connected} focus solely on the stability, while assuming a deterministic performance from the communication network. 
Such an assumption is not practical for platoons that coexist with 5G cellular networks, since interference from uncoordinated cochannel transmissions by other users, vehicles, and platoons can substantially impact the system's performance. 
Clearly,  despite the interdependent performance of communication and control systems in a platoon, there is a lack in existing works that jointly study the wireless and control system performance for vehicular platoons.

The main contribution of this paper is a novel, integrated control system and V2V wireless communication network framework for autonomous vehicular platoons. 
In particular, we first analyze the stability of the control system in a platoon, and, then, we determine the maximum tolerable transmission delay to maintain the stability of the platoon.
Next, we use stochastic geometry and queuing theory to perform end-to-end delay analysis for the V2V communication link between two consecutive vehicles in the platoon. 
Based on the maximum wireless system delay and the theoretical end-to-end delay, we conduct reliability analysis for the wireless communication network.
Here, reliability is defined as the possibility of the wireless system meeting the maximum delay requirements from the control system. 
Finally, we optimize the design of the control system to improve the reliability of the communication network.
\textit{To our best knowledge, this is the first work that considers both control system and V2V wireless communication network for a wireless connected autonomous platoon.}
The novelty of this work lies in the following key contributions:

\begin{itemize}
	\item We propose an integrated control system and V2V wireless communication performance analysis framework to guarantee the overall operation of wireless connected vehicular platoons. 
	In particular, we analyze two types of control system stability, plant stability and string stability, for the platoon, and derive the maximum wireless system delay that guarantees both types of stability. 
	We then consider a highway model that models the distribution of platoon vehicles and non-platoon vehicles and derive the complementary cumulative density function (CCDF) for the signal-to-interference-plus-noise-ratio (SINR) of V2V communication links.
	Given the derived CCDF expression, we make use of queuing theory to determine the end-to-end wireless system delay, including queuing, processing, and transmission delay for the V2V link in the platoon.  	  
	\item We use the derived delay to study how the wireless network can meet the control system's delay needs. 
	In particular, we derive a lower bound on the wireless network reliability (in terms of delay) needed to guarantee plant and string stability.
	In addition, we find an approximated reliability expression for vehicular network scenarios in which the wireless system delay is dominated by the transmission delay.
	\item We propose two optimization mechanisms to effectively design the control system so as to improve the reliability of the wireless network. 
	In particular, we use the dual update method to find the sub-optimal control system parameters that increase the lower bound and the approximated values of the wireless network reliability.
	\item Simulation results corroborate the stability and SINR analysis and validate the effectiveness of the proposed joint control and communication framework.
	%analyze system performance. 
	The results show how key parameters, such as the distribution density of non-platoon vehicles, packet size, spacing between two platoon vehicles, and  platoon size, affect the reliability of a platoon.
	The results also show that, by optimizing the control system, the approximated reliability and the reliability lower bound of the wireless network can increase by as much as $15$\% and $15$\%, respectively.  
\end{itemize}

%\label{SectionSystemModel}

Combining the theoretical analysis and simulation results, we obtain important guidelines for the joint design of the wireless network and the control system in vehicular platoons.
In particular, system parameters, such as the number of followers, the bandwidth, and the spacing between two consecutive vehicles in the platoon should be properly selected to ensure the stability of the control system. 
Meanwhile, the control system should also be optimized to relax the delay constraints of the wireless network, thereby improving the reliability of the platoon system.

The rest of the paper is organized as follows. 
Section \uppercase\expandafter{\romannumeral2} presents the system model. 
In Section \uppercase\expandafter{\romannumeral3}, we perform stability analysis for the autonomous platoon. 
The end-to-end delay and reliability analysis are presented in Section \uppercase\expandafter{\romannumeral4}. Section \uppercase\expandafter{\romannumeral5} shows how to optimize the design of the control system. 
Section \uppercase\expandafter{\romannumeral6} provides the simulation results, and conclusions are drawn in Section \uppercase\expandafter{\romannumeral7}.

\vspace{-0.15in}
\section{System Model}
\vspace{-0.05in}
In this section, we first discuss the highway traffic model for the platoon, the channel model for the vehicular communication network, and the vehicles' control model.
We will then perform an interference analysis for vehicles within the platoon and introduce the joint communication and control problem for wireless autonomous vehicular platoons.
\vspace{-0.2in}
\label{SectionII}
\subsection{Highway Traffic Model}\vspace{-0.05in}

\begin{figure}[!t]
	\centering
	\includegraphics[width=5.5in,height=2.3in]{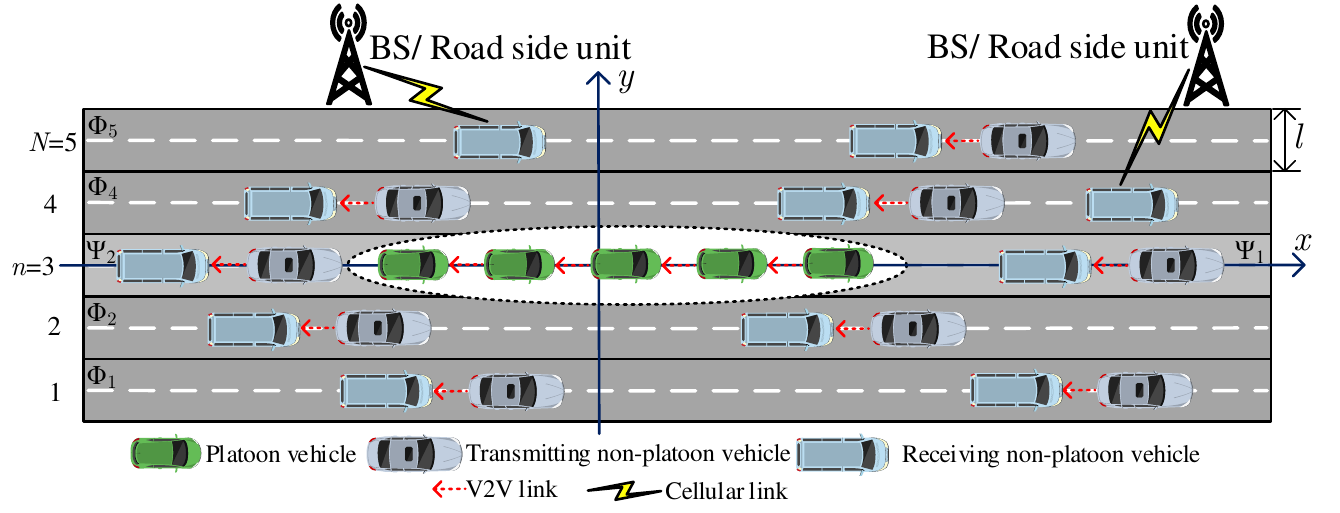}
	\vspace{-0.2in}
	\DeclareGraphicsExtensions.
	\caption{A highway traffic model where vehicles in the dashed ellipse operate in a platoon and other vehicles out of the platoon drive individually.}
	\label{systemmodel}
	\vspace{-0.5in}
\end{figure}

Consider a highway traffic model that is composed of a number of autonomous vehicles driving in a platoon and multiple vehicles driving individually, as shown in Fig. \ref{systemmodel}.
All vehicles (inside and outside the platoon) communicate with one another using V2V communications. 
Vehicles can also exchange information with nearby road side units or cellular base stations (BSs) via vehicle-to-infrastructure (V2I) communications. 
Although both the IEEE 802.11p standard and cellular vehicle-to-everything (C-V2X) communications are widely considered for vehicular networking, here, we focus on the C-V2X communication due to its enhanced performance (e.g., extended communication range and achieve a higher reliability) compared with the IEEE 802.11p standard \cite{lucero2016cellular}.
%Vehicles, out of the platoon, can function as transmitter that send information to other vehicles or play as receivers that receive data from other vehicles via vehicle-to-vehicle (V2V) links or the base station (BS) via cellular links.
Each lane in the highway model has the same width $l$, and vehicles are considered to travel along the horizontal axes in these lanes. 
As shown in Fig. \ref{systemmodel}, we label all $N$ lanes based on their relative locations and we assume that platoon vehicles are moving in platoon lane $n$ while the other $N-1$ lanes are designated as non-platoon lanes.
%Moreover, vehicles are assumed to drive in the same direction for the ease of presentation in Fig. \ref{systemmodel}, but in reality, the traffic flow for each lane can go into the opposite direction.
%Based on the relative locations, we denote the lane with platoon vehicles as platoon lane, and other $N-1$ lanes are considered as non-platoon lanes. 
Accordingly, we define the set $\Phi$ of transmitting vehicles driving on non-platoon lanes and the set $\Psi$ of transmitting non-platoon vehicles on the platoon lane.
In particular, $\Phi$ consists of multiple subsets $\Phi_{h}$, $h\in \{1,2,...,n-1,n+1,...,N\}$, of transmitting non-platoon vehicles on non-platoon lane $h$.
%$N-1$ and $N_{\text{bottom}}\geq 0$ as the number of lanes in the groups of top and bottom lanes, and characterize the transmitting non-platoon vehicles in top lanes, center lane, and bottom lanes as the sets $\phi$, $\varphi$, and $\psi$, respectively.
%That is, $\phi$ and $\psi$ are the unions of vehicles in each lane in the top and bottom groups, i.e., $\phi = \phi_{1} \cup ... \cup \phi_{N_{\text{top}}}$, and $\psi = \psi_{1} \cup ... \cup \psi_{N_{\text{bottom}}}$.
%Note that sets $\phi$ and $\psi$ consist of transmitting non-platoon vehicles in each top lane and bottom lane, respectively, which can be expressed subsets $\phi_{i}$, $0 \leq i \leq N_{\text{top}}$, and $\psi_{j}$, $0 \leq j \leq N_{\text{bottom}}$.
However, set $\Psi$ is only composed of two subsets of transmitting vehicles: subset $\Psi_{1}$ of vehicles that drive ahead of the platoon and subset $\Psi_{2}$ that includes vehicles moving behind the platoon.
%A summary of notations is listed in Table \ref{tableParameters}.

When considering a highway model in a large area, we can leverage spatial point processes from stochastic geometry to capture the distribution of transmitting non-platoon vehicles in the highway \cite{5621983}.
In particular, we characterize the distribution of transmitting vehicles on non-platoon lane $h \neq n$, as a homogeneous Poisson point process (PPP) with density $\lambda_{h}$. 
Moreover, for the platoon lane, the transmitting vehicles driving ahead of the platoon follow a nonhomogeneous PPP \cite{gong2014interference} where the distribution density for vehicles ahead of the platoon is $\lambda_{n}^{(1)}$ and the density for points elsewhere is $0$.
Similarly, the transmitting vehicles driving behind the platoon also follow a nonhomogeneous PPP where the distribution density of points behind the platoon is $\lambda_{n}^{(2)}$ and the density for points elsewhere is $0$.
%we assume that locations of non-platoon transmitters follow one-dimensional Poisson point process (PPP) with different densities.
%In particular, we characterize the distribution density $\lambda_{i}^{\text{top}}$, $1 \leq i \leq N_{\text{top}}$, for vehicles in the $i$th top lane, and $\lambda_{j}^{\text{bottom}}$, $1 \leq j \leq N_{\text{bottom}}$, for vehicles in the $j$th bottom lanes with $\lambda_{0}^{\text{top}}, \lambda_{0}^{\text{bottom}}=0$.
%Moreover, for the center lane, we assume the distribution density of transmitting non-platoon vehicles ahead of the platoon as $\lambda_{1}^{\text{center}}$ and distribution density of transmitting non-platoon vehicles behind the platoon as $\lambda_{2}^{\text{center}}$.
%Similar to \cite{papadimitratos2009vehicular}, there is only a portion of vehicles in each lanes will participate into vehicle-to-vehicle (V2V) communications and functions as transmitters.
%According to the thinning theorem of PPP \cite{haenggi2012stochastic}, we assume the portion as $0 \leq \beta \leq 1$, which is same for each lane, and therefore the density of transmitters for each lane can be the product of $\beta$ and the corresponding vehicle distribution density.
Furthermore, as shown in Fig. \ref{systemmodel}, we consider a Cartesian coordinate system centered on the rear bumper of an arbitrarily selected platoon vehicle.
In this case, the rear bumper location of each vehicle in the highway model can be mathematically represented using the coordinate system.
For example, for a vehicle that is sharing the same lane with the platoon, the location can be expressed as $(x,0)$, where $x$ is the signed distance to the centered vehicle.
However, for vehicle driving on lane different from the platoon lane, the corresponding location is $(x,(h-n)l)$, where $x$ captures the signed distance to the $y$ axis, and $h$ denotes the lane number.

\begin{figure}[!t]
	\centering
	\includegraphics[width=4.4in,height=1.2in]{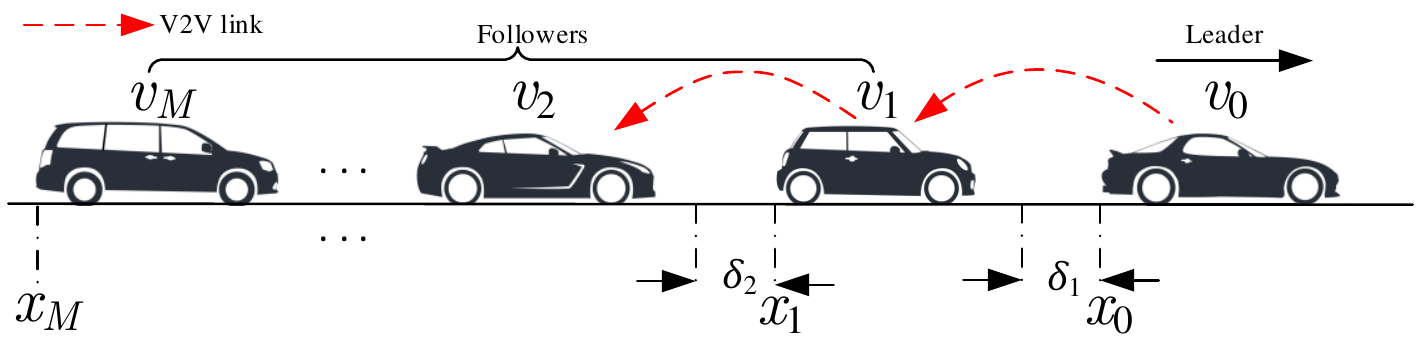}
	\vspace{-0.5cm}
	\DeclareGraphicsExtensions.
	\caption{Leader-follower model: a vehicular platoon with one leader and $M$ followers. }
	\label{systemmodel1}
	\vspace{-0.45in}
\end{figure}

For vehicles inside the platoon, we consider a leader-follower model \cite{jin2013stability}, as shown in Fig. \ref{systemmodel1}.
The leader-follower model is composed of a set $\mathcal{M}$ of $M\!+\!1$ cars where the leading vehicle is the leader and the remaining $M$ vehicles are the followers.
%In the platoon, there are $N$ followers and one leader. In essence, the first vehicle is the leader and other vehicles are the followers.
The location of each vehicle in the platoon is captured by the rear bumper position $(x_{i},0), i\!\in\!\mathcal{M}$, and it can be measured by sensors, like the global positioning system (GPS).
%Also, to help the autonomous vehicle increase awareness of the environment, V2V communications will be used to exchange a variety of information such as critical information (e.g., location, velocity, HD maps), and non-critical information (e.g., videos) inside the platoon.  
In addition, each following vehicle can communicate with the preceding vehicle via a V2V link to collect information, such as the target speed and spacing of the platoon and current speed and location of the preceding vehicle. 
By using the collected information, following vehicles can coordinate their movements to keep a target distance to the corresponding preceding vehicles and form the vehicular platoon.
The need for such information collection in platoons has been validated via many real-world models \cite{bergenhem2012overview} and \cite{liu2001effects}.

%\begin{table}[!t]
%	%	\normalsize
%	\small
%	\begin{center}
%		%\centering
%		\caption{\small Summary of notations.}
%		\label{tableParameters}
%		\resizebox{14cm}{!}{
%			\begin{tabular}{|c|c|}
%				\hline
%				\textbf{Parameter} & \textbf{Description} \\  \hline
%				$l$ & Width of each lane \\ \hline
%				$N$, $n$ & Number of lanes and the label of the platoon lane  \\ \hline
%				$\Phi$, $\Psi$  & Sets of transmitting non-platoon vehicle driving on the non-platoon lanes and platoon lane   \\  \hline
%				$\lambda_{i}, \lambda_{n}^{1}, \lambda_{n}^{2}$  & PPP distribution densities of vehicles    \\ \hline
%				$\mathcal{M}$, $M$  & Set of platoon vehicles and number of followers in the platoon  \\ \hline
%				${\color{blue}P_{t}}$ & Transmission power of vehicle  \\ \hline
%				$g^{j}_{i-1,i}$ & Channel gain of subcarrier $j$ for the channel between vehicle $i-1$ and $i$ \\ \hline
%				$d_{i-1,i}$, $d_{\text{dense}}$, $d_{\text{sparse}}$ & Distance between vehicle $i-1$ and $i$, and for sparse and dense traffic  \\ \hline
%				$m$  & Nakagami channel parameter  \\ \hline
%				$\alpha$  & Path loss exponent  \\ \hline
%				$\omega$  & Total cellular bandwidth  \\ \hline
%				$\sigma^{2}$ & Power of noise    \\ \hline
%				$\hat{L}$, $\hat{v}$ & Target distance and velocity  \\ \hline
%				$v_{\text{max}}$ & Maximum driving speed  \\ \hline
%				$a_{i},b_{i}$ & Control parameters of vehicle $i$  \\ \hline
%				$S$ & Packet size  \\ \hline
%		\end{tabular}}
%	\end{center}
%\end{table}

\vspace{-0.2in}
\subsection{Channel Model and Interference Analysis}
\vspace{-0.05in}
Here, we assume that V2I and V2V transmissions are managed over orthogonal frequency resources, and, therefore, there is no interference between V2I and V2V links \cite{yaacoub2015qoe}. 
%Similar to the work in , we assume that V2I and V2V communications use different frequency resource.
In particular, for V2V communications, we assume a V2V underlay network using which the available cellular bandwidth $\omega$ is reused by all V2V links outside the platoon. 
Meanwhile, we consider an orthogonal frequency-division multiple access (OFDMA) scheme in which the BS will allocate orthogonal subcarriers to platoon vehicles so as to avoid interference between concurrent V2V transmissions inside the platoon\footnote{Due to the small velocity difference between platoon vehicles and the use of sub-6 GHz frequencies, effect of the Doppler shift can be neglected (or assumed to be handled via proper Doppler shift mitigation techniques) \cite{6757685}.}. 
Such an allocation is possible given that the typical platoon will not have a very large number of vehicles and, hence, will require only a few subcarriers.
%In this case, the V2V links can coexist simultaneously without experiencing interference from other links in the \emph{same} platoon.
However, due to bandwidth sharing with non-platoon vehicles, the followers will encounter interference from other V2V links outside the platoon.
According to the channel measurement results presented in \cite{he2014vehicle} and \cite{cheng2007mobile}, we model the V2V channels in the platoon as independent Nakagami channels with parameter $m$  to characterize a wide range of fading environments for V2V links.
%Also, another reason is that Nakagami channel is a more refined fading model that can represent Ricean, Rayleigh fading models Ricean, Rayleigh, or
%fading that is more severe than Rayleigh distribution. 
Also, in general, Nakagami channel models can describe a wide range of fading environment of vehicular networks \cite{cheng2007mobile}.
%the V2V propagation model can be perfectly modeled Nakagami fading channel, and thus, we consider a Nakagami-m fading channel for V2V communications.
Therefore, in the platoon, the received power at any follower $i \in \mathcal{M}$  from the transmission of platoon car $i\!\!-\!\!1$ by using subcarrier $j$ is $P_{i,j}(t)\!=\!P_{t} g_{i,j}(t) (d_{i-1,i}(t))^{-\alpha}$,
where $P_{t}$ is the transmit power of each vehicle, $g_{i,j}(t)$ follows a Gamma distribution with shape parameter $m$, $d_{i-1,i}(t)$ is the distance between vehicles $i-1$ and $i$ inside the platoon, and $\alpha$ is the path loss exponent.
Since line-of-sight signals from non-platoon vehicles to the platoon vehicles do not always exist, we model these channels as independent Rayleigh fading channels \cite{acosta2006doubly}.
Consequently, the overall interference at car $i$ is the sum of two following interference terms:
\begin{align}
\label{non_platoon}
&I_{i}^{\text{non-platoon}}(t) = \sum_{j_{1}=1}^{n-1} \sum_{c \in \Phi_{j_{1}}}
P_{t} g_{c,i}(t)(d_{c,i}(t))^{-\alpha} +
\sum_{j_{2}=n+1}^{N} \sum_{c \in \Phi_{j_{2}}}
P_{t} g_{c,i}(t)(d_{c,i}(t))^{-\alpha}, \\ 
\label{platoon}
&I_{i}^{\text{platoon}}(t) = \sum_{j_{3}=1}^{2} \sum_{c \in \Psi_{j_{3}}}
P_{t} g_{c,i}(t)(d_{c,i}(t))^{-\alpha},
\end{align}
where $j_1 \in [1,...,n-1], j_2 \in [n+1,...,N], j_3 \in [1,2]$, $d_{c,i}(t)$ denotes the distance between vehicles $c$ and $i$, and $g_{c,i}(t)$ refers to the channel gain from vehicle $c$ to $i$ at time $t$, which follows an exponential distribution. 
%The total interference is thereby $I_{i}(t)=I_{i}^{\text{non-platoon}}(t)+I_{i}^{\text{platoon}}(t)$.
Using (\ref{non_platoon}) and (\ref{platoon}), the SINR of the V2V link on subcarrier $j$ from car $i-1$ to $i$ will be:
\begin{align}
\label{sinr}
\gamma_{i,j}(t) = \frac{P_{i,j}(t)}{I_{i}^{\text{non-platoon}}(t)+I_{i}^{\text{platoon}}(t)+\sigma^{2}},
\end{align}
where $\sigma^{2}$ is the variance of Gaussian noise. Using (\ref{sinr}), we can obtain the throughput of the V2V link between vehicles $i-1$ and $i$ as
$R_{i,j}(t) = \omega_{j}\log_{2}(1+\gamma_{i,j}(t))$, where $\omega_{j}$ is the bandwidth of subcarrier $j$. 
Here, we assume that the bandwidth for each subcarrier is $\frac{\omega}{M}$.
%For the ease of exposition, when doing interference and delay analysis for the wireless system in the following sections, we will omit the time-varying factor $(t)$ hereafter. To simplify our notations, we remove the time variable $t$, hereafter. 

\vspace{-0.2in}
\subsection{Control Model}
\vspace{-0.05in}
%\begin{figure}[!t]
%	\centering
%	\includegraphics[width=5.5in,height=1.400in]{SystemModel2.pdf}
%	\DeclareGraphicsExtensions.
%	\vspace{-0.3cm}
%	\caption{Basic structure of a platoon system. }
%	\label{systemmode2}
%	\vspace{-0.8cm}
%\end{figure}
To realize the target spacing  for the platoon, the CACC system in each vehicle will brake or accelerate according to the difference between the actual distance and target spacing slot to the preceding vehicle.
That is, if the difference is positive, the vehicle must speed up so that the distance to the preceding car meets the platoon's target spacing. Otherwise, the vehicle must slow down.
This distance difference is defined as the spacing error $\delta_{i}(t)$:
\begin{equation}
\label{spacingerror}
\delta_{i}(t) = x_{i-1}(t) - x_{i}(t) - \hat{L}(t),  i \in \mathcal{M},
\end{equation}
where $\hat{L}(t)$ is the target spacing at time $t$ for the platoon. The distance difference, $d_{i-1,i}(t) = x_{i-1}(t) - x_{i}(t)$, is also commonly known as the \emph{headway}.
In addition, the velocity error will be:
\begin{equation}
\label{velocityerror}
z_{i}(t) = v_{i}(t) - \hat{v}(t),
\end{equation}
where $v_{i}(t)$ represents the velocity of vehicle $i$ at time $t$, and $\hat{v}(t)$ is the target velocity at time $t$ for the platoon.
Similar to \cite{jin2013stability}, we assume that the spacing requirement $\hat{L}(t)$ and velocity requirement $\hat{v}(t)$ are constants at time $t$. 
Also, similar to the optimal velocity model (OVM) introduced in \cite{bando1995dynamical}, to realize the stability of a platoon system, the acceleration or deceleration of each vehicle must be determined by two components:
1) the difference between headway-dependent and actual velocities, and 2) the velocity difference between a given vehicle and its preceding vehicle.
%As shown in Fig. \ref{systemmode2}, different from velocity data transferred via V2V communications, we assume that the headway information can be measured by the radar sensor without  time delay.
Hence, to guarantee that both velocity and spacing errors converge to zero, we use the following control law to determine the acceleration $u_{i}(t)$ of vehicle $i$ \cite{bando1995dynamical}
\begin{equation}
\label{controlLaw}
u_{i}(t)=a_{i}[V(d_{i-1,i}(t-\tau_{i-1,i}(t)))-v_{i}(t)]+ b_{i}[v_{i-1}(t-\tau_{i-1,i}(t) )-v_{i}(t)],
\end{equation}
where $\tau_{i-1,i}(t)$ captures the wireless system delay between vehicle $i$ and its preceding vehicle, $a_{i}$ is the associated gain of vehicle $i$ for the difference of the headway-dependent velocity
and the actual speed, and $b_{i}$ is the associated gain of vehicle $i$ for the velocity difference between cars $i-1$ and $i$. 
Here, we assume that, for each V2V link, the transmitter will also transmit a timestamp when the information is sent. Thus, the receiving vehicle can measure the transmission delay $\tau_{i-1,i}(t)$ and calculate $d_{i-1,i}(t-\tau_{i-1,i}(t))$.
Also, note that the associated gains $a_{i}$ and $b_{i}$ essentially capture the sensibility of the CACC system to respond to changes of the distance and velocity.
Moreover, the headway-dependent velocity $V(d)$ should satisfy following properties:
1) in dense traffic, the vehicle will stop, i.e., $V(d)= 0$ for $d\leq d_{\text{dense}}$,
2) in sparse traffic, the vehicle can travel with its maximum speed $V(d)= v_{\text{max}}$, which is also called free-flow speed, for $d \geq d_{\text{sparse}}$, and
3) when $d_{\text{dense}}<d<d_{\text{sparse}}$, $V(d)$ is a monotonically increasing function of $d$.
We define the function $V(d)$ \cite{jin2013stability}:
\begin{equation}
\label{Vhfunction}
V(d) = \begin{cases}
0, &\text{if} \hspace{1mm} d < d_{\text{dense}}, \\
v_{\text{max}} \times \left(\frac{d - d_{\text{dense}}}{d_{\text{sparse}}-d_{\text{dense}}}\right),  &\text{if} \hspace{1mm} d_{\text{dense}}\leq d\leq d_{\text{sparse}}, \\
v_{\text{max}}, &\text{if} \hspace{1mm} d_{\text{sparse}}< d. \\
\end{cases}
\end{equation}

%For this model, our goal is to design a control law that is equivalent to finding the control parameters,  $a_{i}$ and $b_{i}$, $i \in \mathcal{N}$ so that all followers can drive with the target velocity $\hat{v}$, which is the speed of the leader, and maintain the target inter-vehicle distance $\hat{L}$. Note that $\hat{v}=V(\hat{L})$.3r 

To guarantee the stable operation of the platoon system, it is important to jointly consider the communication and control systems for the platoon.
In particular, on the one hand, for a given control system setup, one can design the wireless network so as to meet the delay requirement of V2V links and prevent the instability of the control system. 
%one can find the delay requirements for V2V communications that can prevent the instability of the control system.
On the other hand, given the state of the wireless system, one can also optimize the design of the control system to relax the delay requirements for the communication system.
In the following sections, we will first conduct stability analysis for the control system and find the wireless system delay requirements to realize the stable operation of the control system.
Then, based on the distribution of vehicles, we derive the CCDF of the SINR of V2V links in the platoon. 
To model the delay, we consider two queues in tandem for the V2V link, and leverage queuing theory to derive the end-to-end delay, including queuing, processing, and transmission delay. 
%we will derive the end-to-end delay, including the queuing, processing, and transmission delay, for the V2V link in the platoon. 
Then, we derive the lower bound and approximated expressions for the wireless system reliability, defined as the probability that the wireless system meets the delay requirements from the control system.
Moreover, we optimize the design for the control system to maximize the derived reliability metrics of the wireless network. 
\vspace{-0.2in}
\section{Stability Analysis of the Control System}
\vspace{-0.05in}
\label{stabilityanalysis}
For the leader-follower platoon model, the inevitable wireless system delay in (\ref{controlLaw}) can negatively impact the stability of the platoon system.
Here, we perform stability analysis for the control system in presence of a wireless system delay.
We analyze two types of stability: plant stability and string stability \cite{jin2013stability}. 
Plant stability focuses on the convergence of error terms related to the inter-vehicle distance and velocity, while string stability pertains to the error propagation across the platoon.
Using this stability analysis, we obtain the wireless system delay thresholds that can ensure plant and string stability for the control system. 
%design guidelines for the allocation of transmit power and spectral resources to vehicles in the platoon's communication system.
%Moreover, based on the communication and control co-design requirements, we characterize the \emph{reliability} of the wireless system, defined as the probability of the wireless system meeting the control system's delay needs. 
\vspace{-0.2in}
\subsection{Plant Stability}
\vspace{-0.05in}
Plant stability requires all followers in a platoon to drive with the same speed as the leader and maintain a target distance from the preceding vehicle.
In other words, plant stability requires both the spacing and speed errors of each vehicle to converge to zero. 
This convergence also requires the first-order derivative of the error terms in (\ref{spacingerror}) and (\ref{velocityerror}) to approach to zero.
Thus, we can take the first-order derivative of (\ref{spacingerror}) and (\ref{velocityerror}) as:
\begin{equation}
\label{deriavativeoftwo}
\begin{cases}
\dot{\delta_{i}}(t)=z_{i-1}(t)-z_{i}(t), \\
\dot{z_{i}}(t)=A_{i}\delta_{i}(t-\tau_{i-1,i}(t))+B_{i} z_{i-1}(t-\tau_{i-1,i}(t)) - C_{i}z_{i}(t), \\
\end{cases}
\end{equation}
where $\dot{\delta_{i}}(t)$ and $\dot{z_{i}}(t)$ are variables differentiated with respect to time $t$, $A_{i}=\frac{a_{i}v_{\text{max}}}{d_{\text{sparse}}-d_{\text{dense}}}$, $B_{i}=b_{i}$, and $C_{i}=a_{i}+b_{i}$.
Note that since the leading vehicle with index 0 always drives with the target velocity, its velocity (spacing) error is $z_{0}(t)=0$ ($\delta_{0}(t)\!\!=\!\!0$).
As observed from (\ref{deriavativeoftwo}), to guarantee the convergence of the derivatives, both $z_{i}(t)$ and $\delta_{i}(t)$ should asymptotically approach to zero. Therefore, the zero convergence of spacing and speed errors is equivalent to the asymptotical zero convergence of their first-order derivatives.

%Also, as the channel gains of different V2V links follow the same distribution and two adjacent vehicles in a platoon are always close to each other, we assume that the time delay $\tau_{i-1,i}(t) = \tau(t), \forall i \in \mathcal{M}$.
Then, after the BS collects spacing and velocity errors for all of the followers, we can find the augmented error state vector $\boldsymbol{e}(t)\!\!=\!\![\delta_{1}(t),\delta_{2}(t),...,\delta_{M}(t),z_{1}(t),z_{2}(t),...,z_{M}(t)]^T$ and obtain
\begin{equation}
\label{error}
\dot{\boldsymbol{e}}(t) = \begin{bmatrix}
\boldsymbol{0}_{M\times M} & \boldsymbol{\Omega}_{1} \\ \boldsymbol{0}_{M\times M} & \boldsymbol{\Omega}_{2}
\end{bmatrix}\boldsymbol{e}(t)+ \sum_{i=1}^{M}\begin{bmatrix}
\boldsymbol{0}_{M\times M} & \boldsymbol{0}_{M \times M} \\ \boldsymbol{\Omega}^{i}_{3} & \boldsymbol{\Omega}^{i}_{4}
\end{bmatrix}\boldsymbol{e}(t-\tau_{i-1,i}(t)),
\end{equation}
where
\begin{align}
&\boldsymbol{\Omega}_{1}=
\begin{bmatrix}
-1 & 0 & 0 &  \dots & 0 & 0 \\
1 & -1 & 0 &\dots & 0 & 0\\
0 & 1 & -1 &\dots & 0 & 0\\
\vdots & \vdots & \vdots & \ddots & \vdots& \vdots\\
0 & 0 & 0 & \dots & 1 & -1
\end{bmatrix}_{M \times M}, \\
&\boldsymbol{\Omega}_{2} = \text{diag}\{-C_{1}, -C_{2}...,-C_{M}\}_{M \times M}, 
\end{align}
and the elements in $\boldsymbol{\Omega}^{i}_{3} \in \R^{M \times M}$ and $\boldsymbol{\Omega}^{i}_{4}\in \R^{M \times M}$ are defined as 
\begin{equation}
[\boldsymbol{\Omega}^{i}_{3}]_{m_1,m_2} =  \begin{cases}
A_{i}, &\text{if} \hspace{1mm} m_1 = m_2 = i, \\
0, &\text{otherwise},
\end{cases}  \hspace{0.5mm}
[\boldsymbol{\Omega}^{i}_{4}]_{m_1,m_2} = \begin{cases}
B_{i}, &\text{if} \hspace{1mm} m_1 = i, m_2 = i-1, i > 1, \\
0, &\text{otherwise}.
\end{cases} 
\end{equation}
For ease of presentation, we rewrite $\boldsymbol{M}_{1}\!\!=\!\! \begin{bmatrix}
\boldsymbol{0}_{M\times M} & \boldsymbol{\Omega}_{1} \\ \boldsymbol{0}_{M\times M} & \boldsymbol{\Omega}_{2}
\end{bmatrix}$, $\boldsymbol{M}_{2}^{i}\!\!=\!\! \begin{bmatrix}
\boldsymbol{0}_{M\times M} & \boldsymbol{0}_{M \times M} \\ \boldsymbol{\Omega}^{i}_{3} & \boldsymbol{\Omega}^{i}_{4}
\end{bmatrix}$, and use $\boldsymbol{e}$ to denote $\boldsymbol{e}(t)$ hereinafter.
Since plant stability requires the spacing and velocity errors to converge to zero, the error vector $\boldsymbol{e}=\boldsymbol{0}_{2M \times 1}$ should be asymptotically stable.

To guarantee the plant stability for a platoon, the delay experienced by a V2V link should be below a threshold. 
Next, we derive the delay threshold to guarantee plant stability.
\begin{theorem}
	\label{theorem1}
	The plant stability of the platoon can be guaranteed if $a_{i}^2+b_{i}^2+2a_{i}b_{i}-4a_{i}\geq0$ and the maximum delay of the V2V link between vehicles $i-1$ and $i$, $i\in \mathcal{M}$, in the platoon satisfies:
	\begin{align}
	\label{SINR11}
	&\tau_{i-1,i}(t) \leq  \tau_{1}=\frac{\lambda_{\min}(\boldsymbol{M}_{3})}{\lambda_{\max}\big(\boldsymbol{M}_{4}\big)},
	\end{align}
	where $\boldsymbol{M}_{3}\!=\!-2(\boldsymbol{M}_{1}+\sum_{i=1}^{M} \boldsymbol{M}_{2}^{i})$, $\boldsymbol{M}_{4}\!=\!\sum_{i=1}^{M}(\boldsymbol{M}^{i}_{2}\boldsymbol{M}_{1}\boldsymbol{M}_{1}^T(\boldsymbol{M}^{i}_{2})^T) +\sum_{i=2}^{M}(\boldsymbol{M}^{i}_{2}\boldsymbol{M}^{i-1}_{2}(\boldsymbol{M}^{i-1}_{2})^T\\(\boldsymbol{M}^{i}_{2})^T)+2M k\boldsymbol{I}_{2M\times 2M}$ with $k> 1$, and 
	$\lambda_{\max}(\cdot)$ and $\lambda_{\min}(\cdot)$ represent the maximum and minimum eigenvalues of the corresponding matrix.
	\begin{proof}[Proof:\nopunct]
		Please refer to Appendix \ref{prooffortheorem1}.
	\end{proof}
\end{theorem}
\vspace{-0.15in}
Hence, when $a_{i}^2+b_{i}^2+2a_{i}b_{i}-4a_{i}\geq0$ and $\tau_{i-1,i}(t)\! \leq \!\tau_{1}, i\!\in\! \mathcal{M}$, the following vehicles will eventually drive with the same speed as the leading vehicle and keep an identical distance to the corresponding preceding vehicles.
\vspace{-0.2in}

\subsection{String Stability}
\vspace{-0.05in}
Beyond plant stability, we must ensure that the platoon is \emph{string stable}.
In particular, if the disturbances, in terms of velocity or distance, of preceding vehicles do not amplify along with the platoon, the system can have string stability and the safety of the system can be secured \cite{hall2005vehicle}.
%To analyze string stability, we also consider the wireless delay $\tau_{i-1,i}(t) = \tau(t), \forall i \in \mathcal{M},$ between vehicles $i-1$ and $i$.
%the worst-case scenario in which all V2V links in the platoon experience the maximum time delay $\tau^{(2)}$ due to the wireless channel.
To find the delay requirement guaranteeing string stability, we first obtain the transfer function between two adjacent vehicles by finding the Laplace transform of both equations in (\ref{deriavativeoftwo}):
\begin{equation}
\label{transferfunction}
T_{i}(s) = \frac{z_{i}(s)}{z_{i-1}(s)}= \frac{A_{i}+sB_{i}e^{-s\tau_{i-1,i}(t)}}{s^{2}+C_{i}s+A_{i}}, i\in \mathcal{M}.
\end{equation}
%Then, by using the , $e^{x} \approx \frac{1+0.5x}{1-0.5x}$ , we further simplify (\ref{transferfunction}) by finding an approximation of  in the numerator. 
Here, without loss of generality, we assume that vehicles in the platoon are identical, and, thus, the associated gains are equal, i.e., $a_{i}=a$ and $b_{i}=b$, $i\in \mathcal{M}$.
Based on \cite{swaroop1996string}, we know that string stability is guaranteed as long as the magnitude of the transfer function satisfies $|T_{i}(jf)|\leq1$ for $f\in \mathbb{R}^{+}$, where $f$ represents the frequency of sinusoidal excitation signals generated by the leader.
Thus, by using Pad{\'e} approximation \cite{baker1975essentials} for $e^{-s\tau_{i-1,i}(t)}$ in the numerator of (\ref{transferfunction}), we can find an approximated analytical result of the maximum wireless system delay to satisfy $|T_{i}(jf)|\leq1$, $f\in \mathbb{R}^{+}$, in the following proposition.
\vspace{-0.1in}
\begin{proposition}
	\label{theorem2}
	The string stability of the system in (\ref{controlLaw}) can be guaranteed if $a+2b-2\geq0$, and the maximum delay of the V2V link between vehicles $i-1$ and $i$, $i\in \mathcal{M}$, in the platoon satisfies:
	\begin{equation}
	\label{SINR2}
	\tau_{i-1,i}(t) \leq \tau_{2} = \frac{C^{2}-2A-B^{2}}{2AC},
	\end{equation}
	where $A\!=\!\frac{av_{\text{max}}}{d_{\text{sparse}}\!-\!d_{\text{dense}}}$, $B\!=\!b$, $C\!=\!a\!+\!b$, and $\tau_{2}$ is the approximated communication delay threshold.
	\begin{proof}[Proof:\nopunct]
		Please refer to Appendix \ref{prooffortheorem2}.
	\end{proof}
\end{proposition}
\vspace{-0.15in}

To sum up, if $a+2b-2\geq0$ and $\tau_{i-1,i}(t) \leq \tau_{2}$, $i \in \mathcal{M}$, the spacing error and velocity error will not amplify along the string of vehicles, guaranteeing the platoon's safety.
To guarantee both plant and string stability for a platoon, we must ensure that the delay encountered by the V2V link is such that $\tau_{i-1,i}(t) \leq \min(\tau_{1}, \tau_{2})$, $i \in \mathcal{M}$, and the control gains should satisfy $a+2b-2\geq0$ and $a^2+b^2+2ab-4a\geq0$.\vspace{-0.15in}

\section{End-to-End Latency Analysis of the Wireless Network}
\vspace{-0.05in}
\label{endtoendLatencyanalysis}
\begin{figure}[!t]
	\centering \vspace{-0.05in}
	\subfloat[Transmitter architecture.]{%
		\includegraphics[width=4in,height=1.2in]{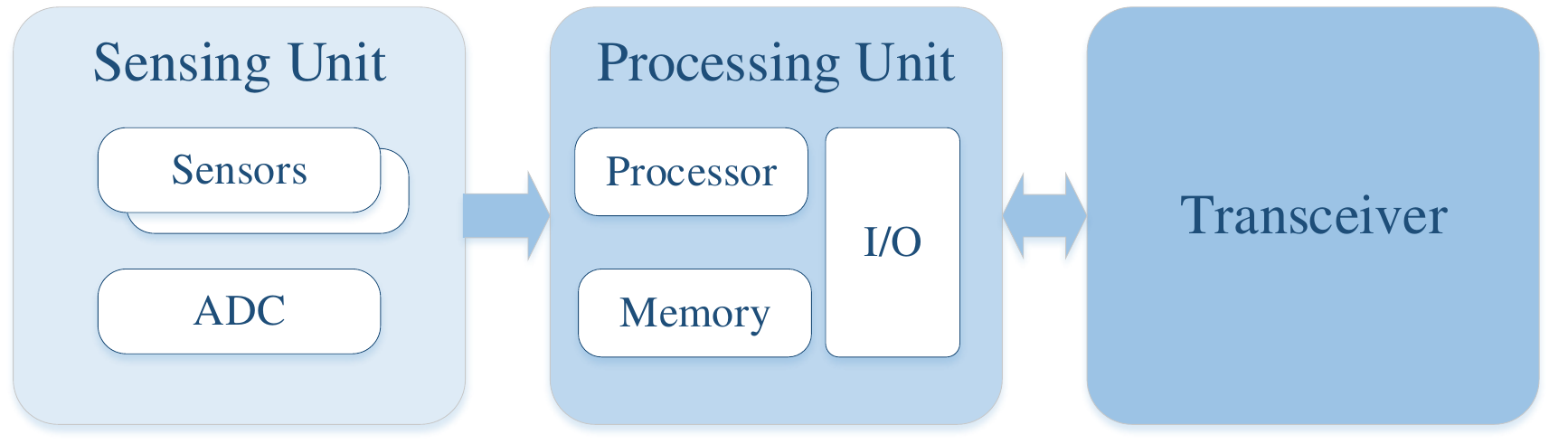}\label{informationPropagation1}
	}	

	\subfloat[Queuing model.]{%
		\includegraphics[width=3.4in,height=0.65in]{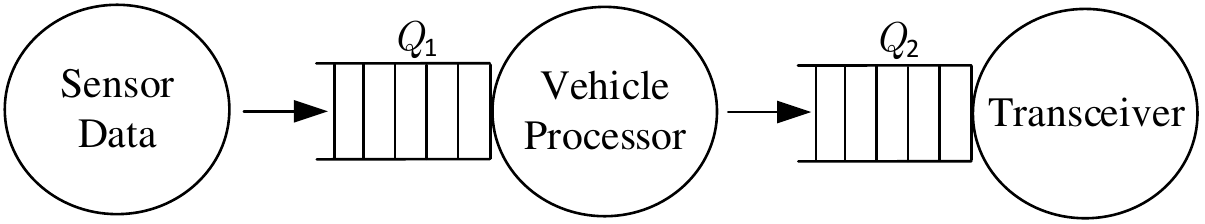}\label{informationPropagation2}
	}	
	\vspace{-0.15in}
	\caption{Data path inside a transmitting vehicle.}
	\label{datapropagation}	
	\vspace{-1cm}
\end{figure}

From the results presented in Section \ref{stabilityanalysis}, to realize the stability for the control system, the wireless V2V network must guarantee that the maximum delay between two consecutive vehicles in the platoon is less than a threshold.
To quantify such wireless system delay, we need to know how a data packet propagates among vehicles as well as the key factors that affect the delay inside the platoon.
As shown in Fig. \ref{datapropagation}\subref{informationPropagation1}, vehicular network information, such as location, will be first collected by sensing units in the vehicle. 
Here, sensing units consist of analog-to-digital converters (ADCs), which convert analog data from the sensor to digital data that can be processed by the processor. 
Then, the processor will not only provide an interface to the sensing unit and the transceiver and execute instructions pertaining to sensing and communication but it is also used to calculate the current speed based on the collected GPS data.
Next, the processor will perform digital to analog conversion and then transmit the analog data to the transceiver so as to transmit to other vehicles via V2V links.
Finally, the receiving vehicles will use the recently received information and sensor data to adjust their acceleration or deceleration based on (\ref{controlLaw}).
Such information exchange is needed since vehicles must be aware of the nearby environment so as to form a target platoon especially under extreme road situations where the velocity and spacing requirements for the platoon can suddenly change and must be exchanged continuously among vehicles in the platoon \cite{bergenhem2012overview,liu2001effects,peng2017resource}.
%know where to go and when to stop \cite{jin2014dynamics}. 
%This process will occur frequently as disturbance always exists because of the ever-changing road condition and the uncertainty of vehicles joining or leaving the platoon. 
To capture the V2V communication delay in the information exchange, we define the queuing model shown in Fig. \ref{datapropagation}\subref{informationPropagation2} (in this model, we assume that sensor information collection and ADC have a negligible delay compared to processing and transmission delay). 
In particular, after being converted at ADCs, each information packet experiences queuing delay and the processing delay at the processor (the first queue $Q_{1}$), and, then, the packet will encounter the queuing delay and the transmission delay at the transceiver (the second queue $Q_{2}$) \cite{melikov2012queuing}.
We define the total time delay from the transmitting vehicle to the receiving vehicle of a V2V link in the platoon as the \emph{end-to-end delay}, including the time spent in the tandem queue, $Q_{1}$ and $Q_{2}$.

%{\color{red}We define such delay from the moment when information is collected at the transmitting vehicle to the time receiving vehicle receives the collected information as end-to-end delay.  
%Also, and thereby the end-to-end delay can be composed of queuing delay and processing delay at the processor of any given vehicle (the first queue $Q_{1}$), and queuing delay and transmission delay at the transceiver (the second queue $Q_{2}$).}

%\begin{figure}[!t]
%	\centering
%	\includegraphics[width=3.3in,height=0.65in]{DataPropagation.pdf}
%	\DeclareGraphicsExtensions.
%	\caption{}
%	\label{datapropagation}
%\end{figure}
\vspace{-0.2in}
\subsection{Queuing Delay and Processing Delay in Queue $Q_{1}$}
\vspace{-0.05in}
Once a vehicle collects the data using its sensors, data needs to be processed locally and then sent to the transceiver.
To model the instantaneous delay $T_{1}$ at the processor, similar to \cite{yang2008towards} and \cite{shuman2006optimal}, we leverage the independence between the sensor measurement and the time interval of two consecutive measurements and consider a Poisson arrival process of the sensor packets with rate $\lambda_{a}$ for the processor.
Also, to track the speed and location changes of preceding vehicles smoothly, we consider that the processor with an infinite-buffer serves the incoming data based on a first come, first served policy \cite{shuman2006optimal}.
Moreover, the service time of the vehicle processor follows an exponential distribution with rate parameter $\mu_{1}>\lambda_{a}$ for guaranteeing the stability of the first queue \cite{kleinrock1976queueing}.
%Similar to the dynamic voltage scaling technique introduced in \cite{puccinelli2005wireless}, the voltage for the processor will dynamically change and the processing rate will change accordingly to realize an efficient power management. 
%To guarantee that the processor will serve packets with a high processing rate when the distance between two consecutive vehicles is small, we assume that the processing rate $\mu_{1}$ can be expressed as a monotonically decreasing function of the distance to the preceding vehicle as follows:
%Consequently, when the distance between two vehicles is small, the service rate $\mu_{1}$ will increase so that the processing time and the queuing delay in queue $Q_{1}$ can be reduced.
%To guarantee the decreasing monotonicity of $\mu_{1}$, 
%\begin{equation}
%\label{Speedfunction}
%\mu_{1}(d) = \left\{ \begin{array}{cc}
%\hspace{4mm} \lambda_{a}, \hspace{23mm}\text{if} \hspace{1mm} d^{\text{radar}}_{\text{max}}<d, \\
%\frac{\lambda_{a}-\mu_{\max}}{d_{\text{max}}^{\text{radar}}}d+ \mu_{\max}, \text{otherwise}, \\
%\end{array} \right.
%\end{equation}
%where $d^{\text{radar}}_{\text{max}}$ is the maximum distance detected by the embedded radar sensor, and $\mu_{\max}$ is the maximum rate of the vehicle's processor.
We assume that each vehicle has only one processor, so the first queue can be modeled as an M/M/1 queue. Thus, according to \cite{kleinrock1976queueing}, the average queuing delay of a packet at the vehicle's processor can be expressed as 
$
\bar{T}_{1}^{q} = \frac{\lambda_{a}}{\mu_{1}(\mu_{1}-\lambda_{a})}.
$
The mean processing time of each packet at the processor can be captured by $\bar{T}_{1}^{s}=1/\mu_{1}$. Based on $\bar{T}_{1}^{q}$ and $\bar{T}_{1}^{s}$, we can obtain the average delay for each packet at the first queue $Q_{1}$:  
\begin{align}
\label{latencyProcessor}
\bar{T}_{1} = \bar{T}_{1}^{q} + \bar{T}_{1}^{s} = \frac{\lambda_{a}}{\mu_{1}(\mu_{1}-\lambda_{a})} + \frac{1}{\mu_{1}}.
\end{align}

\vspace{-0.2in}
\subsection{Queuing Delay and Transmission Delay in Queue $Q_{2}$}
\vspace{-0.05in}
In queue $Q_{2}$, the processing rate of the transceiver is determined by the channel quality and, whenever the buffer is not empty, any incoming packet will have to wait in the buffer.
%After reaching queue $Q_{2}$, the packet needs to wait if the buffer is not empty. Meanwhile, 
According to the Burke's theorem \cite{burke1956output}, when the service rate is bigger than the arrival rate for an M/M/1 queue, the departure process can be modeled as a Poisson process with the same rate.
In this case, given that $\mu_1> \lambda_{a}$ is always satisfied in the first queue $Q_1$, the incoming packet for the second queue $Q_{2}$ follows a Poisson process with rate $\lambda_{a}$.
In addition, we assume an infinite buffer size and a first come, first served policy for $Q_{2}$ \cite{shuman2006optimal}.
Furthermore, the service rate in the second queue $Q_{2}$ is essentially the V2V data rate which will follow a general distribution because of the uncertainty of the wireless channel.   
To characterize such channel uncertainty, we make use of stochastic geometry to analyze the V2V communication performance.
%In particular, provided that the SINR at the receiver is larger than a certain threshold $\theta$, the packet at the head of the queue can be successfully decoded; otherwise, the packet will still stay at the original place and wait for retransmission.
%To obtain the probability of SINR being larger than $\theta$, we will leverage stochastic geometry.
In particular, we assume that the rear bumper of platoon vehicle $i$ is located at the origin of the Cartesian system.
As explained in Section \ref{SectionII}, vehicle $i$ will experience interference from transmitting non-platoon vehicles on any lane.
Next, we take vehicle $i$ as an example and use the Laplace transforms of the experienced interference generated by non-platoon vehicles in the following lemmas.
\vspace{-0.1in}
\begin{lemma}
	\label{lemmaInterference1}
	For an arbitrary vehicle $i$ in the platoon, the Laplace transform of the interference $I_{i}^{\emph{non-platoon}}(t)$ from transmitting vehicles on the non-platoon lanes in (\ref{non_platoon}) can be given by:
	\begin{align}
	\label{lemmaInterference1Expression}
	\mathcal{L}_{i}^{\emph{non-platoon}}(s)=&\prod_{j_{1}=1}^{n-1}\exp\left[-\lambda_{j_{1}}\int_{(n-j_{1})l}^{\infty}\left(1-\frac{1}{1+sP_{t}r^{-\alpha}}\right)\frac{2r}{\sqrt{r^{2}-(n-j_{1})^{2}l^{2}}}dr\right] \nonumber \\
	&\times\prod_{j_{2}=n+1}^{N}\exp\left[-\lambda_{j_{2}}\int_{(j_{2}-n)l}^{\infty}\left(1-\frac{1}{1+sP_{t}r^{-\alpha}}\right)\frac{2r}{\sqrt{r^{2}-(j_{2}-n)^{2}l^{2}}}dr\right].
	\end{align}
	\begin{proof}[Proof:\nopunct]
		Please refer to Appendix \ref{lemmaInterference1proof}.
	\end{proof}\vspace{-0.2in}
\end{lemma}

\begin{lemma}
	\label{lemmaInterference2}
	For an arbitrary vehicle $i$ in the platoon, the Laplace transform of the interference $I_{i}^{\emph{platoon}}(t)$ from transmitting non-platoon vehicles on the platoon lane in (\ref{platoon}) can be given by:
	\begin{align}
	\label{lemmaInterference2Expression}
	\mathcal{L}_{i}^{\emph{platoon}}(s)=\exp\left[-\lambda_{n}^{(1)}\int_{d_{i}^{\emph{head}}}^{\infty} \left(1-\frac{1}{1+sP_{t}r^{-\alpha}}\right)dr -\lambda_{n}^{(2)}\int_{d_{i}^{\emph{tail}}}^{\infty} \left(1-\frac{1}{1+sP_{t}r^{-\alpha}}\right)dr\right],
	\end{align}
	where $d_{i}^{\emph{head}}=x_{0}-x_{i}$ and $d_{i}^{\emph{tail}}=x_{i}-x_{M}$ are the distance from vehicle $i$ to the head and the tail of the platoon, respectively.
	\begin{proof}[Proof:\nopunct]
		The proof is similar to Appendix \ref{lemmaInterference1proof}. However, for vehicles driving on the platoon lane, the distance is directly equal to the horizontal distance.
	\end{proof}
\end{lemma}\vspace{-0.15in}

%\begin{lemma}
%	\label{lemmaInterference3}
%	For vehicle $i$ in the platoon, the Laplace transform of the interference from vehicles on the bottom lanes can be expressed as
%	\begin{align}
%	\mathcal{L}_{i}^{\emph{bottom}}(s)=\prod_{j=1}^{N_{\emph{bottom}}}\exp\left[-\lamb%da_{j}^{\emph{bottom}}\int_{jw}^{\infty} \left(1-\frac{1}{1+s{\color{blue}P_{t}}r^{-\alpha}}\right)\frac{2r}{\sqrt{r^{2}-w^{2}j^{2}}}dr\right].
%	\end{align}
%	\begin{proof}[Proof:\nopunct]
%	\end{proof}
%\end{lemma}

Based on the Laplace transforms of interference terms in (\ref{lemmaInterference1Expression}) and (\ref{lemmaInterference2Expression}), we can obtain the expressions of the mean and variance of the service time $D$ for a single packet as follow. 
\vspace{-0.1in}
\begin{theorem}
	\label{meanandvariance}
	For a single packet transmitted from vehicle $i-1$ to vehicle $i$ in the platoon, the mean and variance of the service time $D$ can be expressed as
	\begin{align}
		\mathbb{E}(D) &=   \int_{0}^{\infty} \frac{SM}{\omega \log_{2}(1+\theta)}f(\theta)d\theta, \label{meanofservicetime}\\
		\emph{Var}(D) &=  \int_{0}^{\infty} \frac{S^{2}M^{2}}{\omega^{2}(\log_{2}(1+\theta))^{2}} f(\theta)d\theta  - \left(\int_{0}^{\infty} \frac{SM}{\omega \log_{2}(1+\theta)}f(\theta)d\theta \right)^{2} \label{varianceofservicetime},
	\end{align}
	where the notation $\E(.)$ represents the mean, $S$ is the packet size in bits, and $f(\theta) = -\frac{d\mathbb{F}(\theta)}{d\theta}$ with
	\begin{align}
	\label{CCDFSINR}
		\mathbb{F}(\theta) = \mathbb{P}(\gamma_{i,j} > \theta)=& \sum_{k=1}^{\beta}(-1)^{k+1}{{\beta}\choose{k}}\exp\left(\frac{-k\eta \theta d_{i-1,i}^{\alpha}}{P_{t}}\sigma^{2}\right)\mathcal{L}_{i}^{\emph{non-platoon}}\left(\frac{k\eta \theta d_{i-1,i}^{\alpha}}{P_{t}}\right)\nonumber \\ &\mathcal{L}_{i}^{\emph{platoon}}\left(\frac{k\eta  \theta d_{i-1,i}^{\alpha}}{P_{t}}\right),
	\end{align}
	and $\eta=\beta(\beta!)^{-\frac{1}{\beta}}$.
	\begin{proof}[Proof:\nopunct]
		Please refer to Appendix \ref{SuccessfullyTransmissionProof}.
	\end{proof}
\end{theorem}\vspace{-0.15in}

Given the distribution of incoming packets and the infinite storage capacity, the second queue can be modeled as an M/G/1 queue. Thus, according to the well-known Pollaczek-Khinchine formula \cite{harrison1992performance}, we can determine the average value of delay $T_{2}$ in the second queue $Q_{2}$, including the transmission delay and the waiting time, as:
\begin{align}
\label{latencyTransceiver}
\bar{T}_{2}= \frac{\rho_{2}+\lambda_{a}\mu_{2}\text{Var}(D)}{2(\mu_{2}-\lambda_{a})}+\mu_{2}^{-1},
\end{align}
where $\mu_{2} = 1/\mathbb{E}(D)$ and $\rho_{2}=\lambda_{a}\mathbb{E}(D)$. 
We assume that the receiving vehicle can be aware of the velocity of the preceding vehicle once it receives the information packet over wireless communications. 
Thus, when platoon vehicles use the received information from V2V communications to coordinate their movements,  the average end-to-end delay of each V2V link can be expressed as $\bar{T}=\bar{T}_{1}+\bar{T}_{2}$.

\vspace{-0.2in}
\subsection{Control-Aware Reliability of the Wireless Network}
\vspace{-0.05in}
To assess the performance of the integrated control and communication system, we introduce a notion of
reliability for the wireless network, defined as the probability $\mathbb{P}(T_{1}\!+\!T_{2}\leq \min(\tau_{1}\!,\!{\tau_{2}}))$ of the instantaneous delay in the wireless network meeting the control system’s delay needs where the notation $\mathbb{P}(.)$ represents the probability.
This reliability measure allows for the characterization of the performance of the wireless network that can guarantee the stability of the platoon's control system. 
Moreover, we will use this deviation to gain insights on the design of wireless networks that can sustain the operation of vehicular platoons. These insights include characterizing
how much transmission power and bandwidth are needed to realize a target reliability.
However, it is challenging to directly derive the probability density functions (PDFs) of the instantaneous wireless network delay. 
The reason is that, in queuing theory, the average waiting time is not derived based on the PDF of the instantaneous waiting time. 
%we cannot derive the average waiting time by using the PDFs of the instantaneous waiting time. 
Instead, the average waiting time is calculated by first deriving the average number of packets staying in the queue and then using Little's law, which is the relationship among the number of packets, the incoming packet rate, and the waiting time \cite{kleinrock1976queueing}. 
As the end-to-end delay is composed of queuing delay, processing delay, and transmission delay, finding the exact PDFs for the instantaneous wireless system delay and the reliability is thereby challenging. 
Alternatively, we will derive a lower bound for the reliability of the wireless network in the following theorem.  
%{\color{red}NEED MORE WORK The lower bound can enable us to obtain the guidelines what is minimum requirement} .
%First of all, we can expresses the communication delay as
%\begin{align}
%\label{CommunicationDelay}
%T = \left\{ \begin{array}{cc}
%\frac{\lambda_{a}}{\mu_{1}(\mu_{1}-\lambda)}+\frac{S}{\mu_{1}}+\frac{\lambda{a}}{\mu_{2}(\mu_{2}-\lambda)}+\frac{S}{\mu_{2}}+\frac{NS}{B\log_{2}(1+\gamma_{i-1,i})}, \text{if} \hspace{1mm} \gamma_{i-1,i} \geq \theta, \\
%\infty ,\hspace{75mm} \text{if} \hspace{1mm} \gamma_{i-1,i} < \theta, \\
%\end{array} \right.
%\end{align}
%where we replace $\mu_{1}(d)$ with $\mu_{1}$.
%Then, the reliability of the wireless system can be obtained in the following theorem.
\begin{theorem}
	\vspace{-0.1in}
	\label{theorem4}
	For the followers in a platoon system, when the average wireless system delay $\bar{T}$ is smaller than the requirement $\min(\tau_{1},\tau_{2})$ of the stability of the control system, a lower bound for the reliability of the wireless network can be given by:
	\begin{align}
	\label{LoweBound1}
	\mathbb{P}(T_{1}\!\!+\!\!T_{2}\leq \min(\tau_{1}\!,\!{\tau_{2}})) &\!\geq\! \max\left(1\!\!-\!\! \frac{\bar{T}_{1}\!\!+\!\!\bar{T}_{2}}{\min(\tau_{1}\!,\!{\tau_{2}})},1\!\!-\!\!
	\exp\left(\bar{T}_{1}\!\!+\!\!\bar{T}_{2}\!\!-\!\!\min(\tau_{1}\!,\!{\tau_{2}}) \ln\left(\frac{\min(\tau_{1}\!,\!{\tau_{2}})}{\bar{T}_{1}\!\!+\!\!\bar{T}_{2}} \right)  \right)\right).
	\end{align}
	\begin{proof}[Proof:\nopunct]
		Please refer to Appendix \ref{proofforTheorem4}.
	\end{proof}\vspace{-0.15in}
\end{theorem}

\begin{corollary}
	\vspace{-0.1in}
\label{corollary1}
	By substituting the delay requirement $\min(\tau_{1}, \tau_{2})$ by $\tau_{1}$ or $\tau_{2}$ in (\ref{LoweBound1}), the lower bounds of the reliability for the wireless network guaranteeing either plant stability or string stability can be obtained.
\end{corollary}\vspace{-0.15in}

Given the lower bounds in Theorem \ref{theorem4} and Corollary \ref{corollary1}, we can deduce key guidelines for the joint wireless network and the control system. 
For instance, to guarantee that the reliability exceeds a threshold, e.g., 95\%, we can ensure that the lower bound in (\ref{LoweBound1}) is equal to the threshold by choosing proper values for the wireless network parameters, such as bandwidth and transmission power.
Meanwhile,
we can increase $\min(\tau_{1},\tau_{2})$ by properly selecting the control parameters, i.e., $a$ and $b$, for the control system to guarantee that the lower bound is equal to the threshold as well.
Moreover, next, we can obtain an approximated reliability expression if the wireless delay is dominated by the transmission delay.  
\vspace{-0.1in}
\begin{corollary}
\label{corollary2}
When the vehicle's processor is highly capable and the incoming packet rate is small, the delay at $Q_{1}$ and the queuing delay at $Q_{2}$ are relatively small compared to the transmission delay at $Q_{2}$. 
In this case, the wireless system delay is dominated by the transmission delay at $Q_{2}$, and the reliability of the wireless network can be thereby approximated by:
\begin{align}
\label{closeForm}
\mathbb{P}(T_{1}\!\!+\!\!T_{2}&\!\!\leq\!\!\min(\tau_{1},{\tau_{2}})) \approx  \sum_{k=1}^{\beta}(-1)^{k+1}{{\beta}\choose{k}}\exp\left(\frac{-k\eta \left(2^{\frac{SM}{\omega \min(\tau_{1},{\tau_{2}})}}\!\!-\!\!1\right) d_{i-1,i}^{\alpha}}{P_{t}}\sigma^{2}\right)\nonumber \\ &\mathcal{L}_{i}^{\emph{non-platoon}}\left(\frac{k\eta\left(2^{\frac{SM}{\omega \min(\tau_{1},{\tau_{2}})}}\!\!-\!\!1\right)d_{i-1,i}^{\alpha}}{P_{t}}\right)\mathcal{L}_{i}^{\emph{platoon}}\left(\frac{k\eta \left(2^{\frac{SM}{\omega \min(\tau_{1},{\tau_{2}})}}\!\!-\!\!1\right) d_{i-1,i}^{\alpha}}{P_{t}}\right).
\end{align}
	\begin{proof}[Proof:\nopunct]
		The proof is analogous to Appendix \ref{SuccessfullyTransmissionProof} and the difference is replacing $\theta$ with $2^{\frac{SM}{\omega \min(\tau_{1},{\tau_{2}})}}-1$ in the CCDF (\ref{CCDFSINR}) of SINR. 
	\end{proof}
\end{corollary}
\vspace{-0.15in}
From Corollary \ref{corollary2}, we can not only infer guidelines for the design of the wireless network and the control system to guarantee a promising reliability, but we can also observe how the interference and noise directly impact the ability of the wireless network to secure the stability of the control system. 
To mitigate such impacts, one needs to develop interference management and noise mitigation mechanisms.
However, when the state of the wireless network is given, we can still guarantee a satisfactory reliability for the platoon system by optimizing the design of the control system, as explained next.\vspace{-0.15in}

\section{Optimized Controller Design}
\vspace{-0.05in}
\label{OptimalControllerDesign}
For a system with fixed control parameters in the control law (\ref{controlLaw}), we can meet the delay requirements in Theorem \ref{theorem1} and Proposition \ref{theorem2} by improving the wireless network performance.
However, when the control parameters are not fixed, we can optimize the design of the control system to relax the constraints on the wireless network without jeopardizing the system stability.
In particular, to improve the reliability of the wireless network, the optimization of the control system can be done depending on the capabilities of the processor and the arrival rate. 
For instance, for scenarios in which the processor is highly capable and the arrival rate is small, we can find control parameters for maximizing $\min(\tau_{1},\tau_{2})$ so as to improve the approximated reliability as per Corollary \ref{corollary2}. 
In contrast, if we consider a general scenario, then we can directly increase the lower bound as per Theorem \ref{theorem4}.
%One way is to  in Corollary , and the other way is to directly maximize the lower bound for the reliability in Theorem . 
%To minimize the delay encountered by the receiving vehicles in the platoon and further increase the reliability, one can increase the assigned bandwidth, raise the transmission power, and move the platoon to a proper location so as to decrease the interference. However, due to the scarcity of bandwidth resource, increasing the assigned bandwidth will always come with a high cost, which the operators are not willing to see.
%Also, raising the transmission power will inevitably introduce a higher interference, leading to the uncertainty of minimizing the delay.
%Moreover, moving the platoon to a proper location needs to perform interaction with other vehicles and do path planing, resulting to a high overhead at the receiving vehicles.
%Instead of making use of the aforementioned methods, one can design the control mechanism so that the delay requirements of guaranteeing the stabilities is not too low and easy for wireless communication system to realize
\vspace{-0.2in}
\subsection{Optimization of the Approximated Reliability }
\vspace{-0.05in}
\label{subsection1}
To improve the reliability of the wireless network in Corollary \ref{corollary2}, we design the control system to maximize the smaller value between the two stability delay requirements, i.e., $\max \min(\tau_{1},\tau_{2})$.
%Also, the smaller value should be bigger than the expectation of the wireless network delay $\bar{T}$, which is the summation of $\bar{T}_{1}$ and $\bar{T}_{2}$.
%Due to the monotonicity of the approximated reliability in Corollary 2 of $\min(\tau_{1},\tau_{2})$, maximizing $\min(\tau_{1},\tau_{2})$ will lead to an increase of the reliability of the wireless network where the wireless delay is dominated by the transmission delay.  
%By maximizing $\min(\tau_{1},\tau_{2})$, we can increase the reliability (\ref{closeForm}) of the wireless network in Corollary \ref{corollary2} when the wireless delay is dominated by the transmission delay, due to the monotonicity of reliability (\ref{closeForm}) in terms of $\min(\tau_{1},\tau_{2})$.
%Here, we assume that the control parameters are the same for each vehicle at time $t$, i.e., ${\color{blue}a_{i}}=a$ and ${\color{blue}b_{i}}=b$, $i\in \mathcal{M}$, 
Here, the optimization problem can be formulated into the following form: 
\begin{align} \label{11}
&\max_{a,b}  \min(\tau_{1},\tau_{2})    \\
&\hspace{0.04in}\text{s.t.} 
%& \hspace{0.28in} \tau \geq \bar{T}_{1} +\bar{T}_{2}, \label{constraint2} \\
\hspace{0.04in} a_{\text{min}}\leq a \leq a_{\text{max}}, b_{\text{min}}\leq b \leq b_{\text{max}} \label{constraint4}, \\
&\hspace{0.27in} a^2+b^2+2ab-4a\geq0, a + 2b -2 \geq 0,\label{constraint41}
\end{align}
where constraint (\ref{constraint4}) guarantees that both control parameters are selected within reasonable ranges, and constraint (\ref{constraint41}) ensures the existence of $\tau_{1}$ and $\tau_{2}$.
%Note that $\tau_{1}$ can be simplified as $\tau_{1}=\frac{\lambda_{\min}(-\boldsymbol{M}_{1}\!-\!\boldsymbol{M}_{2}\!-\!(\boldsymbol{M}_{1}\!+\!\boldsymbol{M}_{2})^T)}{B^{2}(A^2+B^2+C^2)+2k}$ where each element in the matrix $-\boldsymbol{M}_{1}\!-\!\boldsymbol{M}_{2}\!-\!(\boldsymbol{M}_{1}\!+\!\boldsymbol{M}_{2})^T$ can be verified as a linear function in terms of control gains $a$ and $b$. 
Note that, we can obtain $\tau_{2}= \frac{(a+2b)(d_{\text{sparse}}-d_{\text{dense}})-2v_{\text{max}}}{2(a+b)v_{\text{max}}}$ from Proposition \ref{theorem2}. Then, we can replace $\min(\tau_{1},\tau_{2})$ with $\tau$ and the optimization problem in (\ref{11})--(\ref{constraint41}) can be rewritten as following equivalent optimization problem:
%$\tau_{1}=\frac{\lambda_{\min}(-\boldsymbol{M}_{1}\!-\!\boldsymbol{M}_{2}\!-\!(\boldsymbol{M}_{1}\!+\!\boldsymbol{M}_{2})^T)}{\lambda_{\max}(\boldsymbol{M}_{2}\boldsymbol{M}_{1}\boldsymbol{M}_{1}^T\boldsymbol{M}_{2}^T\!+\! \boldsymbol{M}_{2}\boldsymbol{M}_{2}\boldsymbol{M}_{2}^T\boldsymbol{M}_{2}^T\!+\!2k\boldsymbol{I}_{2M\times 2M})}$
%and $\tau_{2}=\frac{C^{2}-2A-B^{2}}{2AB}$, determining whether $\min(\tau_{1},\tau_{2})$ is a convex function in terms of $a$ and $b$ is hard. 
%To find the solution to the optimization problem, we can find an equivalent convex optimization problem in the following lemma.
%The optimization problem in --(\ref{constraint4})	is equivalent to the following convex problem:
\begin{align} \label{newOptim1}
&\max_{a,b,\tau}  \tau   \\
&\hspace{0.04in}\text{s.t.} \hspace{0.04in} \lambda_{\max}(\boldsymbol{M}_{4})\tau - \lambda_{\min}(\boldsymbol{M}_{3}) \leq 0, \label{newOptimCon1} \\
&\hspace{0.28in}  2(a+b)v_{\text{max}}\tau -(a+2b)(d_{\text{sparse}}-d_{\text{dense}}) +2v_{\text{max}}\leq 0, \label{newOptimCon2} \\
&\hspace{0.28in} a^2+b^2+2ab-4a\geq 0, a + 2b -2 \geq 0, \label{newOptimCon3} \\
&\hspace{0.28in}  a_{\text{min}}\leq a \leq a_{\text{max}}, b_{\text{min}}\leq b \leq b_{\text{max}}, \tau> 0,  \label{newOptimCon4}
\end{align}
where constraints (\ref{newOptimCon1}) and (\ref{newOptimCon2}) guarantee that the value of $\tau$ is smaller than the minimum value between $\tau_{1}$ and $\tau_{2}$, constraint (\ref{newOptimCon3}) is analogous to (\ref{constraint41}), and constraint (\ref{newOptimCon4}) ensures that values of  $a$, $b$, and $\tau$ are within reasonable ranges. 
%It can be easily proven that the Hessian matrices of the first four terms on the left-hand side of (\ref{newOptimCon1}) are positive semidefinite, and, thus, the four terms are convex functions in terms of $a$, $b$, and $\tau$. 
%In addition, as each element in the matrix $\left(-\boldsymbol{M}_{1}\!-\!\boldsymbol{M}_{2}\!-\!(\boldsymbol{M}_{1}\!+\!\boldsymbol{M}_{2})^T\right)$ is a linear combination of $a$ and $b$, the minimum eigenvalue of such matrix is a concave function of $a$ and $b$ \cite{boyd2004convex}.
%Hence, the fifth term on the left-hand side of (\ref{newOptimCon1}), i.e., $- \lambda_{\min}\left(-\boldsymbol{M}_{1}\!-\!\boldsymbol{M}_{2}\!-\!(\boldsymbol{M}_{1}\!+\!\boldsymbol{M}_{2})^T\right)$, is also a convex function. 
%Since the summation of convex functions is a convex function, the constraint (\ref{newOptimCon1}) is convex.
%The objective function and other constraints are linear expression in terms of $\tau$, $a$, and $b$. 
%Therefore, the equivalent optimization problem is a convex problem. 

Since the optimization problem in (\ref{newOptim1})--(\ref{newOptimCon4}) is not convex, we use the dual update method, introduced in \cite{yu2006dual}, to obtain an efficient sub-optimal solution.  
%and subgradient decent algorithm \cite{yu2007transmitter}.  
In particular, we iteratively update Lagrange multipliers in the Lagrange function to obtain the optimal values for these Lagrange multipliers, and, then, calculate the optimization variables by solving the dual optimization problem. 
First, we obtain the Lagrange function as \vspace{-0.1in}
\begin{align}
L(v_1,v_2,v_3,v_4)\!=& \tau \!+\! v_{1} ( \lambda_{\min}(\boldsymbol{M}_{3})\!-\!\lambda_{\max}(\boldsymbol{M}_{4})\tau)\!+\!v_{2} ((a\!+\!2b)(d_{\text{sparse}}\!-\!d_{\text{dense}})\!-\!2(a\!+\!b)v_{\text{max}}\tau \!\nonumber \\ &-2v_{\text{max}}) 
+v_{3}(a^2+b^2+2ab-4a)+v_4(a+2b-2), 
\end{align}%
where $v_1,v_2,v_3,v_4$ are the Lagrange multipliers for constraints in (\ref{newOptim1})--(\ref{newOptimCon3}).
Next, we obtain a subgradient of $L(v_1,v_2,v_3,v_4)$ as follows:\vspace{-0.1in}
\begin{align}
&\Delta v_{1}\!\!=\!\! \lambda_{\min}(-\boldsymbol{M}^*_{3}) \!\!-\!\! \lambda_{\max}(-\boldsymbol{M}^*_{4})\tau^{*},
\Delta v_{2}\!\! = \!\! (a^*\!+\!2b^*)(d_{\text{sparse}}\!\!-\!\!d_{\text{dense}})\!-\!2(a^*\!+\!b^*)v_{\text{max}}\tau^* \!-\!2v_{\text{max}}, \label{subgradient1} \\
&\Delta v_{3} = (a^*)^2 + (b^*)^2 +2a^* b^* -4a^*, \Delta v_{4} = (a^*) + (2b^*) -2, \label{subgradient2}
\end{align}
where $\boldsymbol{M}^*_{3}$ and $\boldsymbol{M}^*_{4}$ share the same expression with $\boldsymbol{M}_{3}$ and $\boldsymbol{M}_{4}$ with $a$, $b$, and $\tau$ replaced with the optimal $a^*$, $b^*$, and $\tau^*$.
To prove the subgradients in (\ref{subgradient1}) and (\ref{subgradient2}), we assume $(v_1^{\prime},v_2^{\prime},v_3^{\prime},v_4^{\prime})$ is the updated value of $(v_1,v_2,v_3,v_4)$, and we have \vspace{-0.1in}
\begin{align}
L(v_1^{\prime},v_2^{\prime},&v_3^{\prime},v_4^{\prime}) \!\geq\!  \tau^* \!\!+v_1^{\prime} \Delta v_{1}
+v_2^{\prime} \Delta v_{2} +v_3^{\prime} \Delta v_{3} +v_4^{\prime} \Delta v_{4}  \nonumber \\ &= L(v_1,v_2,v_3,v_4)\!+\!(v_1^{\prime}\!-\!v_{1})\Delta v_{1}\!+\!(v_2^{\prime}\!-\!v_{2})\Delta v_{3}\!+\!(v_3^{\prime}\!-\!v_{3})\Delta v_{3}\!+\!(v_4^{\prime}\!-\!v_4)\Delta v_4.
\end{align}  
Therefore, the results in (\ref{subgradient2}) are proven by using the definition of subgradient. 
After finding the optimal dual variables for $v_1,v_2, v_3, v_4$, we can derive the values of the control gains $a$ and $b$ by solving the dual optimization problem, which is not listed here due to the space limitations.
Moreover, we choose the ellipsoid method to find the dual variables, and all variables will converge in $\mathcal{O}(49log(1/\epsilon))$ iterations where $\epsilon$ is the accuracy \cite{boyd2004convex}.

\vspace{-0.2in}
\subsection{Optimization of the Lower Bound for the Reliability}
\vspace{-0.05in}
To increase the wireless network's reliability derived in Theorem \ref{theorem4}, we can directly maximize the lower bound obtained in (\ref{LoweBound1}) by choosing proper $a$ and $b$. In particular, the optimization function can be formulated as \vspace{-0.1in}
\begin{align}
&\argmax_{a,b} \max\left(1- \frac{\bar{T}_{1}+\bar{T}_{2}}{\min(\tau_{1},{\tau_{2}})},1\!\!-\!\!
\exp\left(\bar{T}_{1}+\bar{T}_{2} - \min(\tau_{1},{\tau_{2}}) \ln\left(\frac{\min(\tau_{1},{\tau_{2}})}{\bar{T}_{1}+\bar{T}_{2}} \right)  \right)\right), \label{optimization2} \\
&\hspace{0.04in}\text{s.t.}\hspace{0.1in} (\ref{constraint4}), (\ref{constraint41}) \nonumber  \\
& \hspace{0.32in} \bar{T}_{1}+\bar{T}_{2} \leq \min(\tau_{1},\tau_{2}), \label{constraint21}
\end{align}\vspace{-0.10in}
where constraint (\ref{constraint21}) is a necessary condition for  calculating the lower bound of the reliability. 
\vspace{-0.3in}

\begin{corollary}
	\label{theorem6}
	The sub-optimal control gains for the optimization problem of the reliability lower bound will be equal to the sub-optimal solution to the convex optimization problem in (\ref{newOptim1})--(\ref{newOptimCon4}) as long as such control parameters can guarantee $\bar{T}_{1}+\bar{T}_{2} \leq \min(\tau_{1},\tau_{2})$.
	%reliability lower bound, the optimal control parameters are the same to the ones selected in Theorem 
%	to meet the following requirements:
%	\begin{align}
%	\label{answer1}
%	&\frac{\partial \max\left(1- \frac{\bar{T}_{1}+\bar{T}_{2}}{\min(\tau_{1},{\tau_{2}})},1\!\!-\!\!
%		\exp\left(\bar{T}_{1}+\bar{T}_{2} - \min(\tau_{1},{\tau_{2}}) \ln\left(\frac{\min(\tau_{1},{\tau_{2}})}{\bar{T}_{1}+\bar{T}_{2}} \right)  \right)\right)}{\partial a} = 0, \\ 
%	&\frac{\partial \max\left(1- \frac{\bar{T}_{1}+\bar{T}_{2}}{\min(\tau_{1},{\tau_{2}})},1\!\!-\!\!
%		\exp\left(\bar{T}_{1}+\bar{T}_{2} - \min(\tau_{1},{\tau_{2}}) \ln\left(\frac{\min(\tau_{1},{\tau_{2}})}{\bar{T}_{1}+\bar{T}_{2}} \right)  \right)\right)}{\partial b}= 0, \label{answer2} \\ \label{answer3}
%	&\bar{T}_{1}+\bar{T}_{2} \leq \min(\tau_{1},\tau_{2}).
%	\end{align}
	\begin{proof}[Proof:\nopunct]	
		Please refer to Appendix \ref{prooffortheorem6}.
	\end{proof}
\end{corollary}
\vspace{-0.15in}

Using the two foregoing optimization problems, we can find appropriate parameters for the control mechanism to improve the performance of wireless networks. 
However, we note that changing the control system parameters may lead to an increase of the manufacturer cost and maintenance spending. 
Nevertheless, the sub-optimal solutions to these optimization problems still provide us with key guidelines on how to modify the control parameters to optimize the platoon's overall operation.

\vspace{-0.2in}
\section{Simulation Results and Analysis}
\vspace{-0.1in}
In this section, we will first validate the theoretical results in Sections \ref{stabilityanalysis} and \ref{endtoendLatencyanalysis} by numerical results. 
Moreover, we present performance analysis for the integrated communication and control system based on the results in Sections \ref{endtoendLatencyanalysis} and \ref{OptimalControllerDesign}.
In particular, we consider a $10$ kilometer-long highway segment with $4$ lanes, and the lane with label $n=4$ is the platoon lane. According to the empirical data collected by the Berkeley Highway Laboratory \cite{BerkeleyHighwayLabWebsite} and its analytical results  \cite{panichpapiboon2013irresponsible}, the density of vehicles on the highway is mostly in the range from $0.01$~vehicle/m to $0.03$~vehicle/m. Therefore, we consider the density of transiting non-platoon vehicles on each lane in the range ($0.005$~vehicle/m, $0.015$~vehicle/m). The values of the parameters used for simulations are summarized in Table \ref{tableParametersValues}. 

\begin{table}[!t]
	\small
	\begin{center}
		\centering
		\caption{\small Simulation parameters.}
		\vspace{-0.1in}
		\label{tableParametersValues}
		\resizebox{16cm}{!}{
			\begin{tabular}{|c|c|c|}
				\hline
				\textbf{Parameter} & \textbf{Description} & \textbf{Value}  \\  \hline
				$l$ & Width of each lane & $3.7$~m %\cite{ioannou2013automated} 
				\\ \hline
				$N$, $n$ & Number of lanes and label of platoon lane & $4$, $4$  \\ \hline
				$P_{t}$ & Transmission power & $27$~dBm  \\ \hline
				$\beta$  & Nakagami parameter &$3$ \cite{he2014vehicle} \\ \hline
				$\alpha$ & Path loss exponent & $3$ \\ \hline
				$\omega$  & Total bandwidth &$40$~MHz  \\ \hline
				$\sigma^{2}$ & Noise variance &$-174$~dBm/Hz   \\ \hline
				$v_{\text{max}}$& Maximum speed  & $30$~m/s \cite{jin2013stability} \\ \hline
				$S$ & Packet size & $3,200$~bits\footnotemark[1], $10,000$~bits \\ \hline
				$M$ & Number of followers  & 6 \\ \hline 
				$d_{\text{sparse}}$, $d_{\text{dense}}$ & Distance for sparse and dense traffic &$35$~m \cite{jin2013stability}, $5$~m \cite{jin2013stability} \\ \hline
				$\lambda_{a}$, $\mu_{1}$ & Incoming rate of packets and maximum processing rate for the processor &$10$~packets/s \cite{ma2009performance}, $10,000$~packets/s \\ \hline
				%$d_{\text{max}}^{\text{radar}}$& Maximum distance detected by the radar sensor  & $200$~m \cite{digikeyRadar}\\ \hline
		\end{tabular}}
	\end{center}
	\vspace{-0.5in}
\end{table}

%\footnotetext[2]The value of  $\lambda_{a}$ is chosen to guarantee enough awareness of speed and location changes of the preceding vehicle.

\vspace{-0.2in}
\subsection{Validation of Theoretical Results}
\vspace{-0.05in}
\begin{figure}[!t]
	\centering
	\subfloat[Plant stability.]{\includegraphics[width=0.47\textwidth]{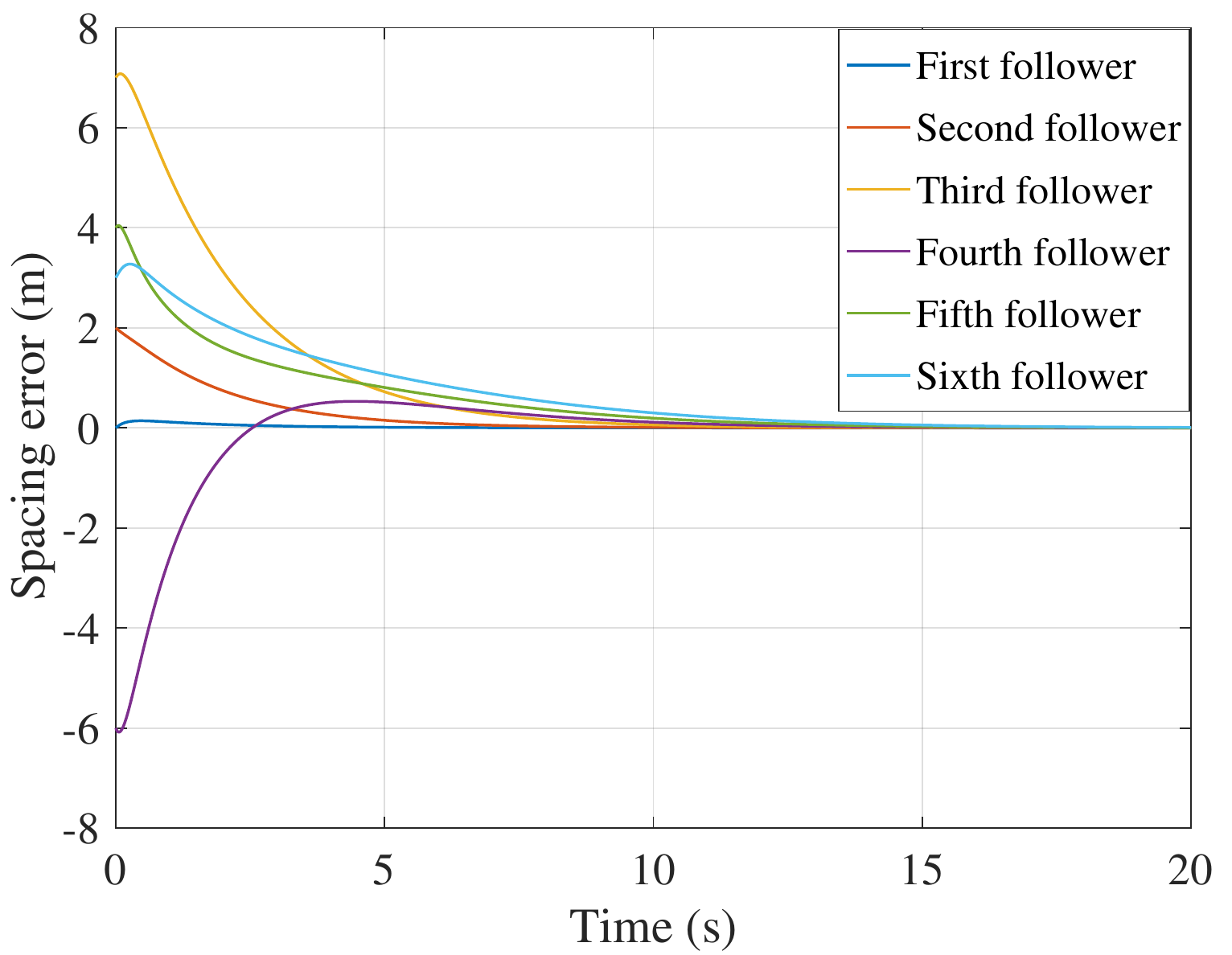}\label{validatePlant}}
	\subfloat[String stability.]{\includegraphics[width=0.47\textwidth]{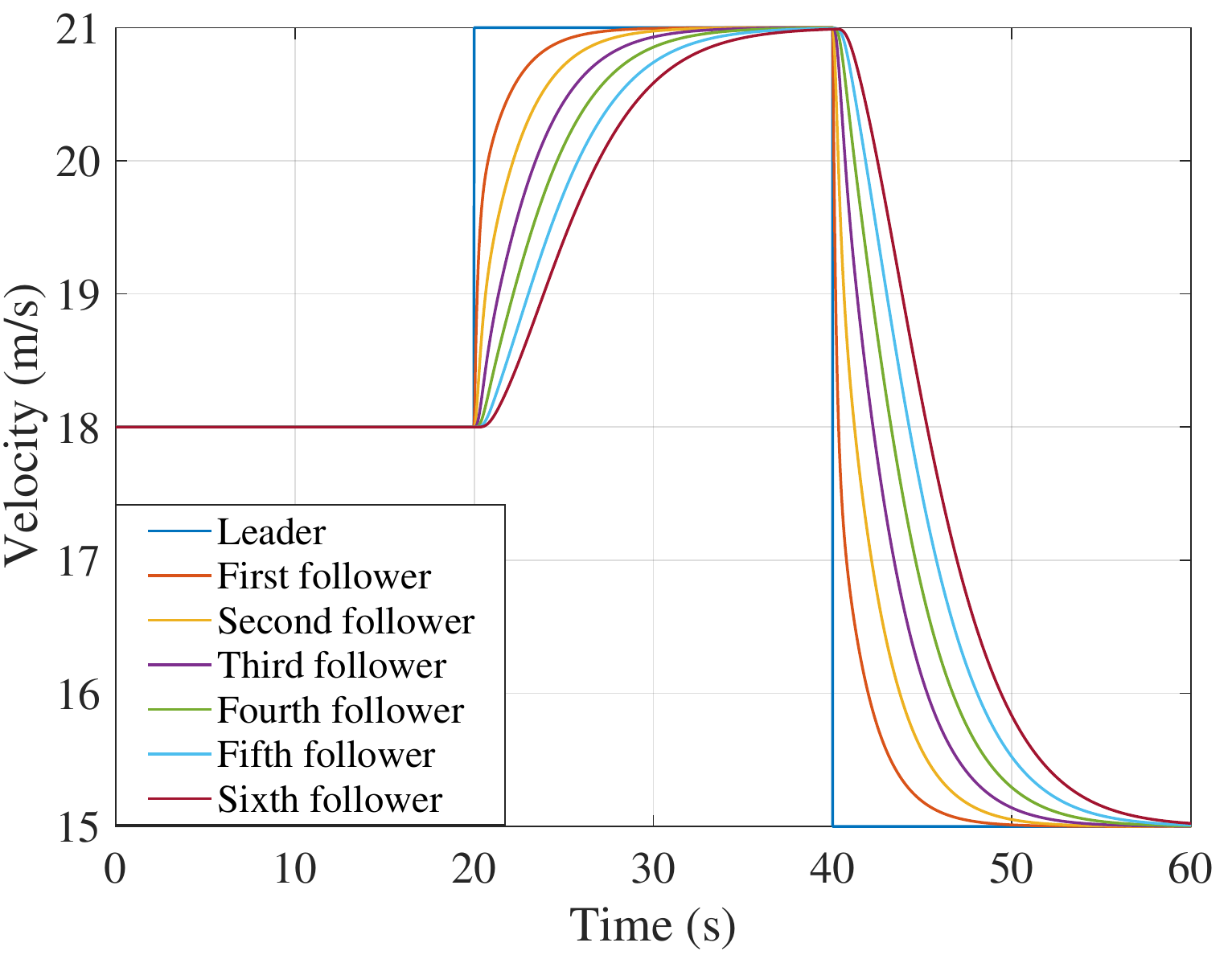}\label{validateString}}
	\hfill
	\vspace{-0.2in}
	\caption{Control system stability analysis validation. }
	\label{stabilityValidation}
	\vspace{-0.45in}
\end{figure}
\footnotetext[1]{The packet size of $S$ is chosen based on the specifications for the Dedicated Short Range Communications (DSRC) safety messages length \cite{kenney2011dedicated}. }

Based on Theorem \ref{theorem1} and Proposition \ref{theorem2}, we can find that the maximum time delay to guarantee the plant stability and string stability are, respectively, $13.9$~ms and $0.5$~s when the control parameters are set to $a=2$ and $b=2$. 
Hence, we first corroborate our analytical results for both types of stability under the minimum of these two delays, i.e., $13.9$~ms.
In particular, we model the uncertainty of the wireless system delay between two adjacent vehicles in the platoon
system as a time-varying delay in the range ($0$, $13.9$~ms).
Vehicles in the platoon are initially assigned different velocities
and different inter-vehicle distances.
Here, the target velocity is $\hat{v}(t)=15$~m/s, and the target inter-vehicle distance is $\hat{L}(t) = 20$~m.

Fig. \ref{stabilityValidation}\subref{validatePlant} shows the time evolution of the spacing errors. 
We can observe that the spacing error will converge to $0$ (a similar
result is observed for the velocity error but is omitted due to space limitations).
Thus, by choosing the maximum time delay derived from Theorem \ref{theorem1} and Proposition \ref{theorem2}, we can ensure the plant stability for the platoon system. 
Next, to verify the string stability, we add disturbances to the leader, that increase the velocity from $18$ to $21$~m/s at $t=20$~s and decrease it from $21$ to
$15$~m/s at $t = 40$~s.
Note that the disturbance might come from changes of road conditions or malfunctions of the control system.
As shown in Fig. \ref{stabilityValidation}\subref{validateString}, the velocity error is not amplified when propagating across the platoon, guaranteeing the string
stability. 
In particular, when the velocity of the leader jumps from $18$ to $21$~m/s, the velocity curve of the sixth follower is more smooth compared with the counterpart of the first follower. 
Clearly, the delay thresholds, found by Theorem \ref{theorem1} and Proposition \ref{theorem2}, can guarantee the plant stability and string stability for the platoon system. 

\begin{figure}[t]
	\centering
	\includegraphics[width=3.3in,height=2.4in]{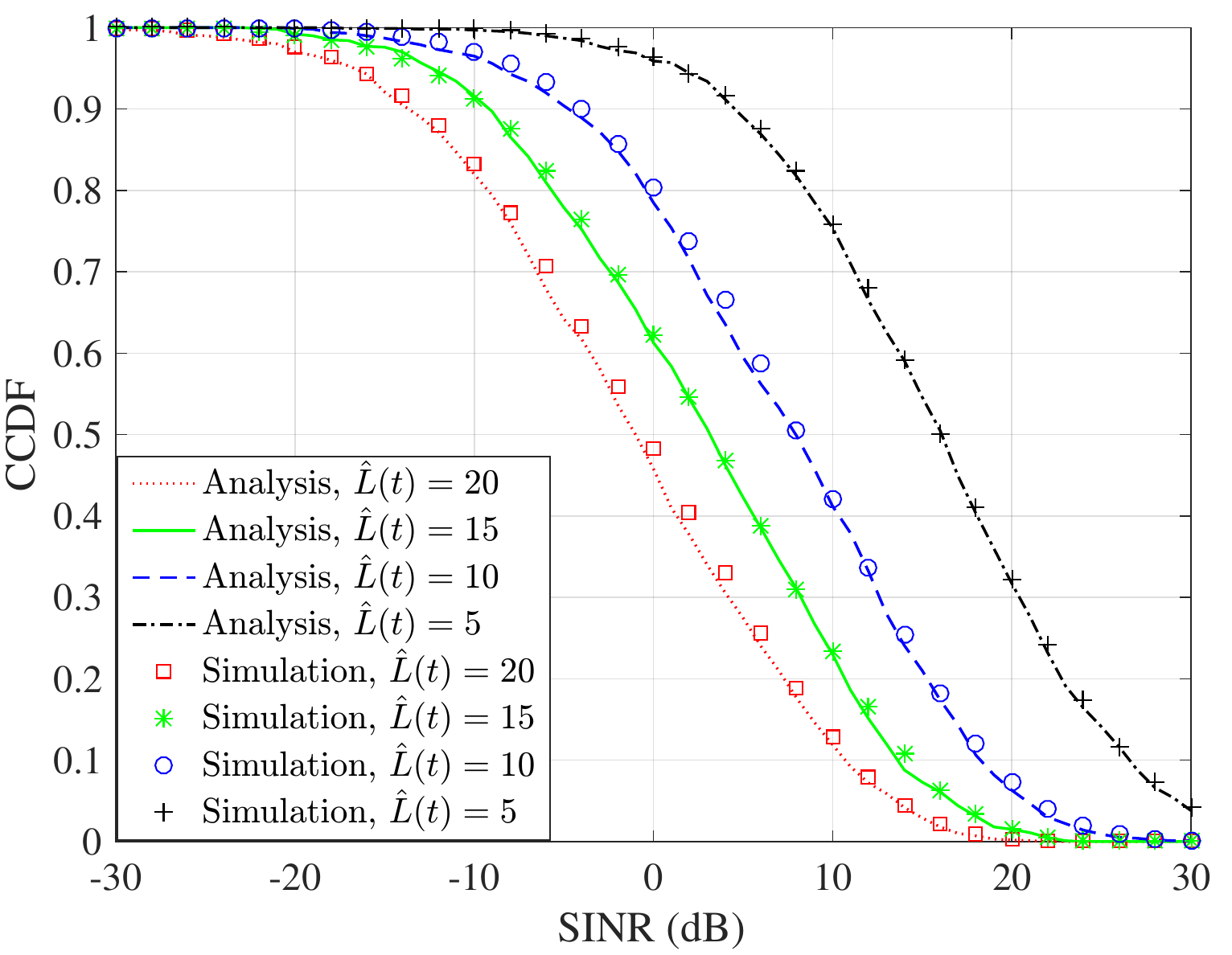}
	\vspace{-0.15in}
	\caption{Validation for the SINR CCDF (\ref{CCDFSINR}) derived in Theorem  \ref{meanandvariance}.}
	\label{validateStochastic}\vspace{-0.3in}
\end{figure}

\begin{figure}
	\centering	
	\begin{minipage}{0.49\textwidth}
		\centering
		\includegraphics[width=1\linewidth]{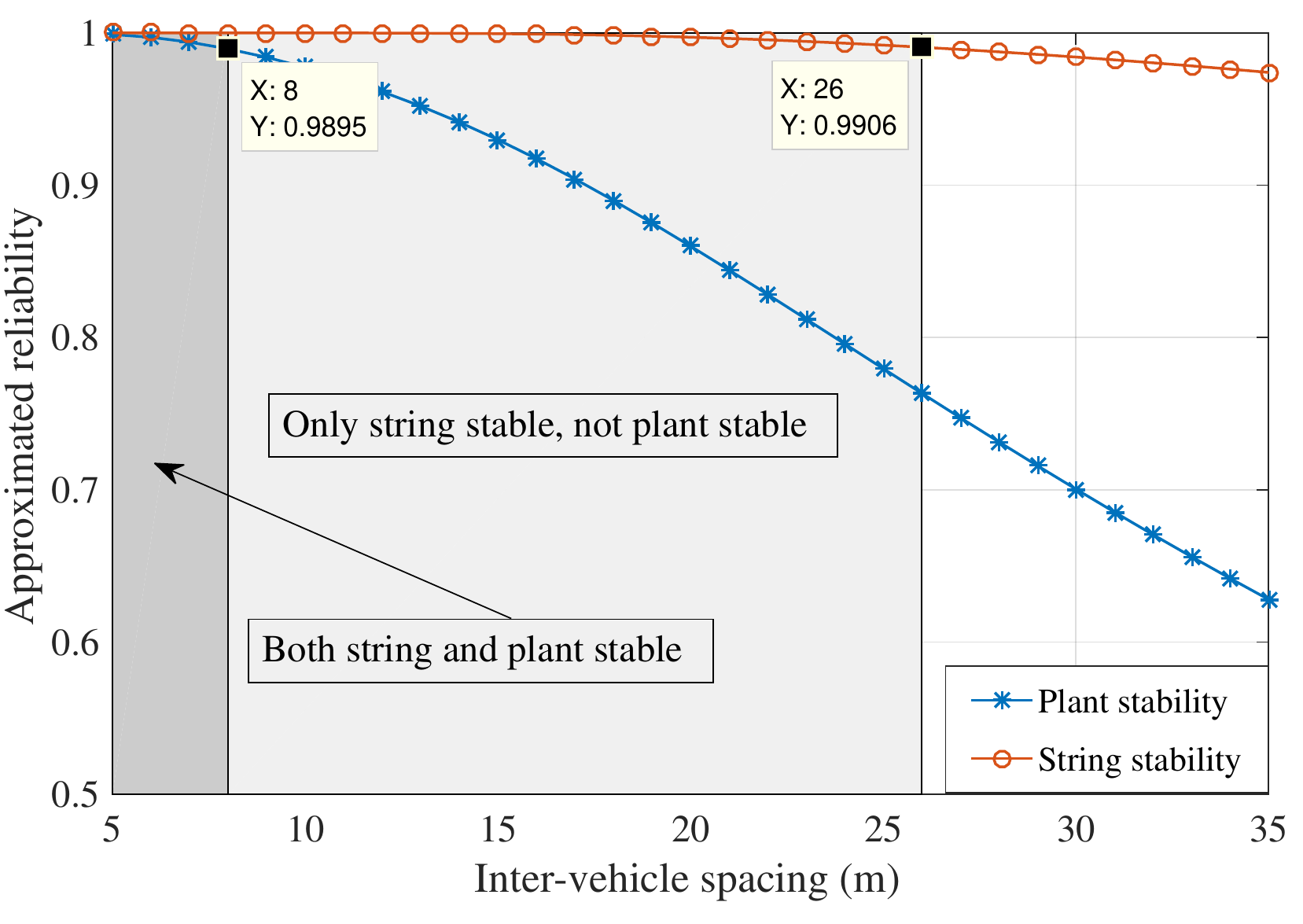}
		\vspace{-0.45in}
		\caption{Approximated reliability performance analysis for platoons with different spacing.}
		\label{reliabilitylowerbound}
	\end{minipage}
	\begin{minipage}{0.49\textwidth}
		\centering
		\includegraphics[width=1\linewidth]{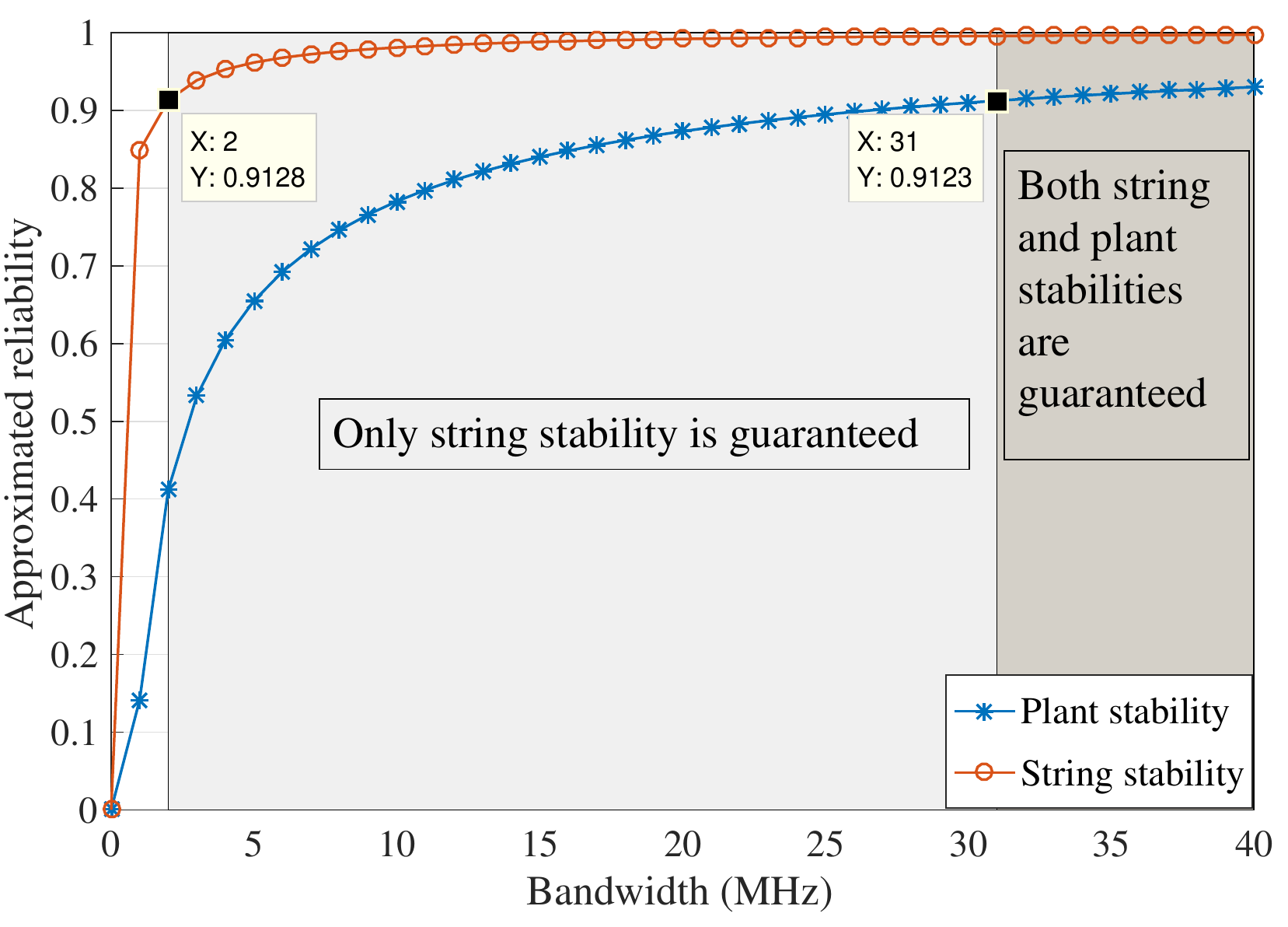}
		\vspace{-0.45in}
		\caption{Approximated reliability for platoon with different total bandwidth.}
		\label{bandwidth}
	\end{minipage}%%%
	\vspace{-0.35in}
\end{figure}

%\begin{figure}[t]
%	\centering
%	\includegraphics[width=3.3in,height=2.4in]{}
%	\vspace{-0.15in}
%	\caption{}
%	\label{}\vspace{-0.4in}
%\end{figure}
%\begin{figure}[!t]
%	\centering
%	\includegraphics[width=3.5in,height=2.8in]{validateStochastic.pdf}
%	\DeclareGraphicsExtensions.
%	\caption{Validation for the CCDF (\ref{CCDFSINR}) of SINR in Theorem \ref{meanandvariance}.}
%	\label{validateStochastic}
%\end{figure}
Fig. \ref{validateStochastic} shows the CCDFs in (\ref{CCDFSINR}) of the SINR derived in Theorem \ref{meanandvariance} for platoons with different spacings between two consecutive platoon vehicles. Here, to characterize the density difference between overtaking lanes and slow lanes, we assume the vehicle density to be $\lambda_{1}=0.01$~vehicle/m, $\lambda_{2}=0.005$~vehicle/m, $\lambda_{3}=0.005$~vehicle/m, $\lambda_{4}^{(1)}=0.01$~vehicle/m, and $\lambda_{4}^{(2)}=0.01$~vehicle/m.
As observed from Fig. \ref{validateStochastic}, the simulation results match the analytical calculations in (\ref{CCDFSINR}), guaranteeing the effectiveness to
derive the mean and variance of the service time based on (\ref{CCDFSINR}) in Theorem \ref{meanandvariance}. 
%of using (\ref{CCDFSINR}) to derive the mean and variance of service time in Theorem \ref{meanandvariance}. 
Moreover, Fig. \ref{validateStochastic} shows that a smaller spacing in the platoon can lead to a higher probability of being at high SINR regions than the one with a larger spacing. 
For example, when $\hat{L}(t)=5$~m, the probability that the SINR will be greater than $10$~dB is around $0.76$, while the counterpart for the platoon with $\hat{L}(t)=15$~m is around $0.24$. 
This is due to the fact that vehicles in the platoon with smaller spacings can receive a signal with higher strength from the vehicle immediately ahead.  
\vspace{-0.2in}

\subsection{Reliability Analysis}
\vspace{-0.05in}

In Figs. \ref{reliabilitylowerbound} and \ref{bandwidth}, we first show the
the approximated reliability in Corollary \ref{corollary2} for the delay requirement obtained from Theorem \ref{theorem1} and Proposition \ref{theorem2} under different inter-vehicle distance and total bandwidth used by the platoon.
%In particular, we consider a scenario with a large size of packets, i.e., $10000$~bits.
As illustrated in Fig. \ref{reliabilitylowerbound}, to guarantee that the approximated reliability of being plant stable exceeds 0.99, the distance between two consecutive platoon vehicles should be smaller than $8$~m. 
However, to reach the approximated reliability 0.99 for string stability, the inter-vehicle spacing should be smaller than $26$~m. 
Moreover, when the inter-vehicle distance is above $26$~m, the platoon cannot achieve an approximated reliability of 0.99 that is needed to ensure string stability or plant stability.
Similarly, as shown in Fig. \ref{bandwidth}, the requirements on bandwidth to achieve the same approximated reliability for ensuing string stability and plant stability are different. 
In particular, to reach the approximated reliability of 0.90 needed for guaranteeing string stability, the total bandwidth $\omega$ is around $2$~MHz. 
In contrast, to guarantee an approximated reliability 0.90 for guaranteeing plant stability, the total bandwidth $\omega$ is approximately $31$~MHz.
Thus, when designing the platoon, we need to properly choose inter-vehicle spacing and bandwidth so as to achieve a target reliability that is needed to guarantee string stability and plant stability. 
\begin{figure}
	\centering	
	\begin{minipage}{0.48\textwidth}
		\centering
		\includegraphics[width=1\linewidth]{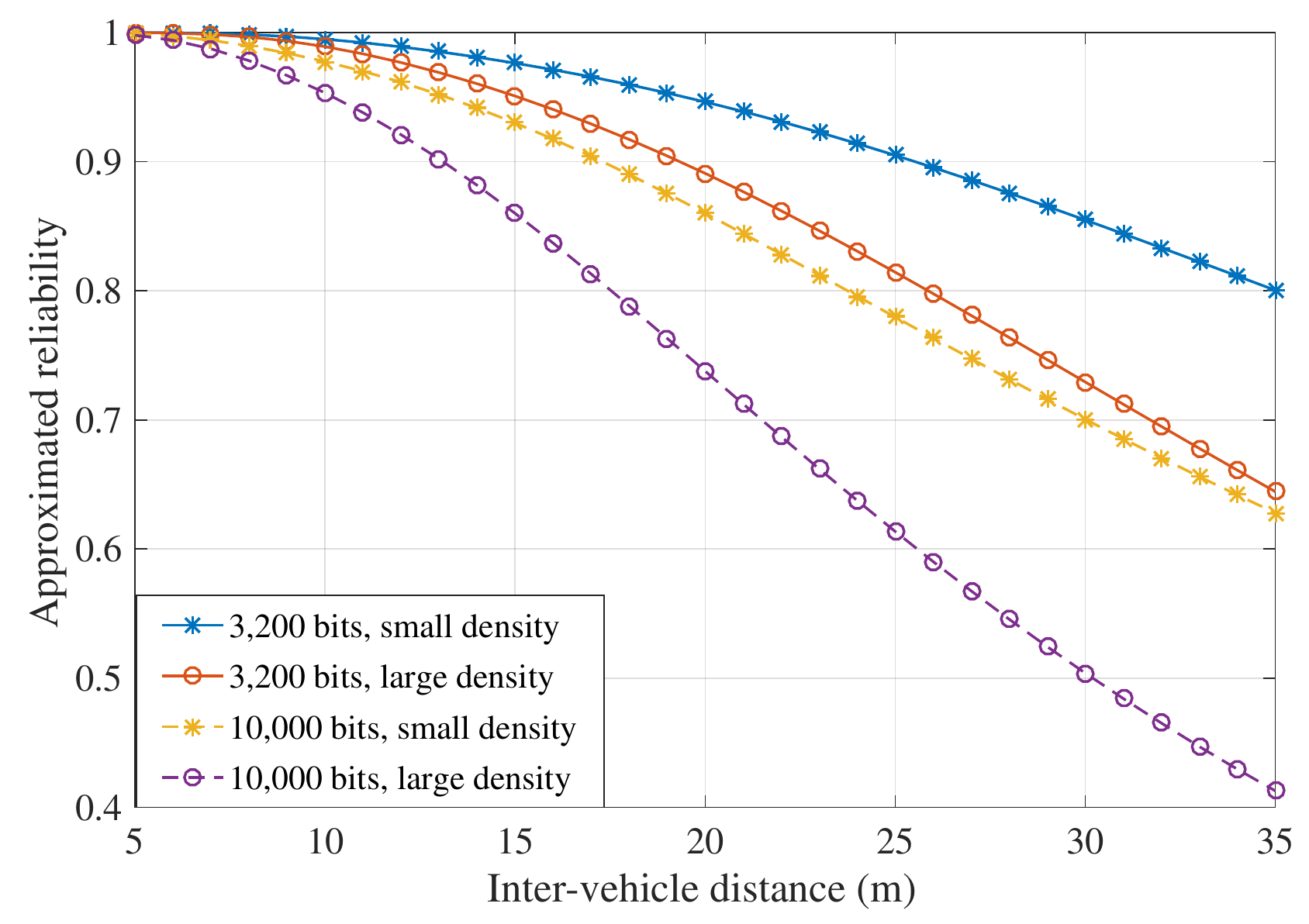}
		\vspace{-0.45in}
		\caption{Approximated reliability for platoons with different density for non-platoon vehicles and  packet sizes.}
		\label{approximatedreliability}
	\end{minipage}
	\begin{minipage}{0.48\textwidth}
		\centering
		\includegraphics[width=1\linewidth]{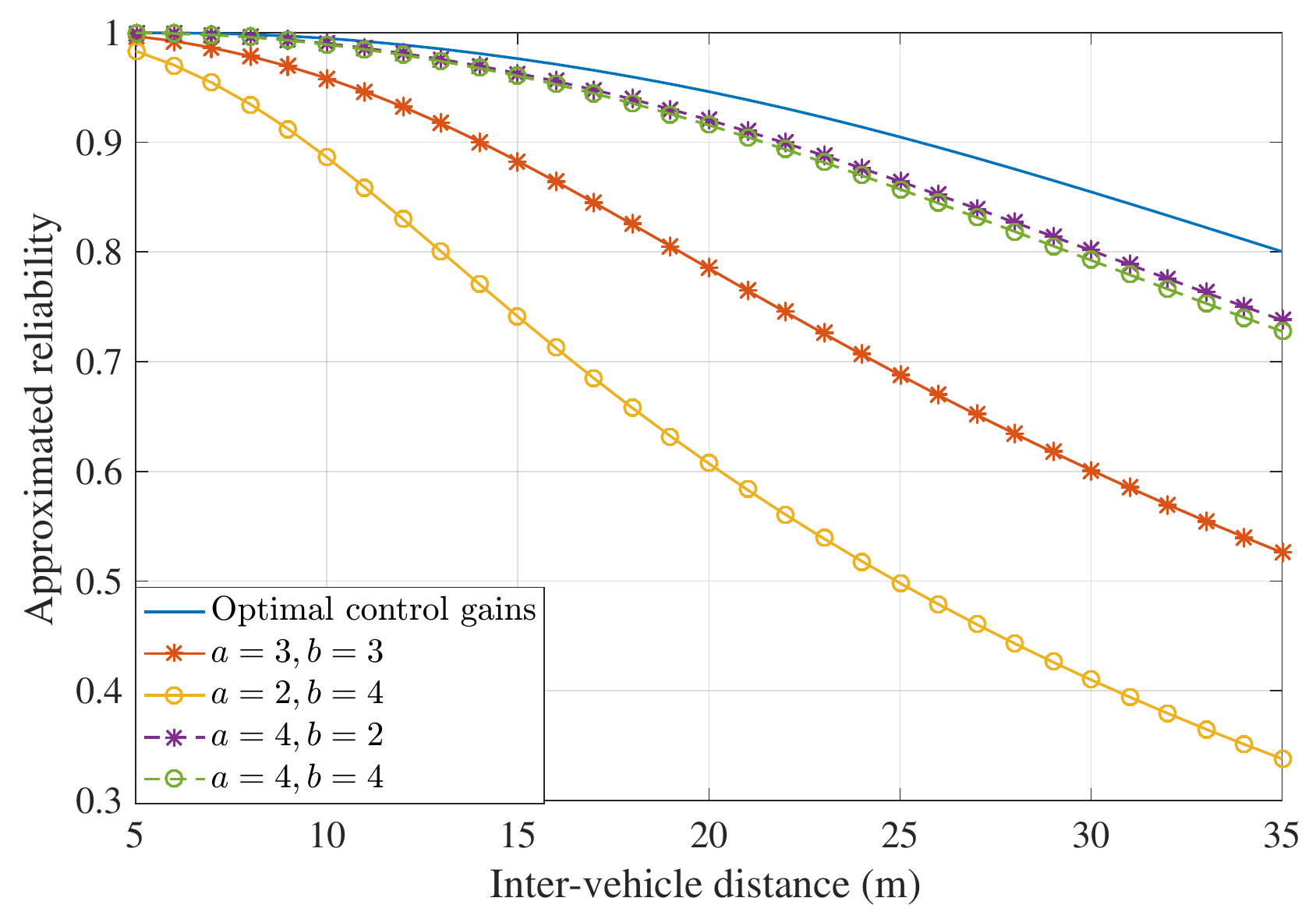}
		\vspace{-0.45in}
		\caption{Approximated reliability for platoons with and without the optimized control system.}
		\label{controlApproximate}
	\end{minipage}
	\vspace{-0.45in}
\end{figure}	
	
Fig. \ref{approximatedreliability} shows the approximated reliability for scenarios with different density of transmitting non-platoon vehicles and different packet sizes. 
We consider two traffic scenarios: the first scenario with small density, i.e.,  $\lambda_{1}\!=\!0.01$~vehicle/m, $\lambda_{2}\!=\!0.005$~vehicle/m, $\lambda_{3}\!=\!0.005$~vehicle/m, $\lambda_{4}^{(1)}\!=\!0.01$~vehicle/m, and $\lambda_{4}^{(2)}\!=\!0.01$~vehicle/m and the second scenario with high density, i.e., $\lambda_{1}\!=\!0.015$~vehicle/m, $\lambda_{2}\!=\!0.01$~vehicle/m, $\lambda_{3}\!=\!0.01$~vehicle/m, $\lambda_{4}^{(1)}\!=\!0.015$~vehicle/m, and $\lambda_{4}^{(2)}\!=\!0.015$~vehicle/m. 
Moreover, we consider two packet sizes, one with $3,200$~bits and the other with $10,000$~bits. 
As observed from Fig. \ref{approximatedreliability}, the approximated reliability decreases with the increase of the distance between two consecutive platoon vehicles and the distribution density of transmitting non-platoon vehicles. 
This is due to the fact that, as the distance or density increases, the SINR will decrease, leading to a decline in data rate and an increase of transmission delay.  
Also, in Fig. \ref{approximatedreliability}, a larger size of packets will increase the transmission time and degrade the reliability. 
%Also, a larger density of transmitting non-platoon vehicle and a large size of transmitting packet will lead to the decrease of the reliability. 
In addition, from Fig. \ref{approximatedreliability}, we can obtain design guidelines on target spacing between two nearby platoon vehicles. 
That is, in order to ensure that the approximated reliability of the wireless network exceeds the target threshold, the platoon spacing should be below a typical value. 
For example, in a scenario with small density of transmitting non-platoon vehicles, the target distance should not be larger than $25$~m so that the approximated reliability can be no less than $0.9$ when transmitting small packets. 
Furthermore, since the target spacing is correlated with the target velocity as shown in (\ref{Vhfunction}), we can also have insights about how to choose the target velocity for the platoon system.

%the platoon distance should not be bigger than $10$~m for guaranteeing the reliability lower bound to be no less than $90$\% in a scenario with lower density of transmitting non-platoon vehicles.  

%\begin{figure}[!t]
%	\centering
%	\includegraphics[width=3.5in,height=2.8in]{approximatedReliability.pdf}
%	\DeclareGraphicsExtensions.
%	\caption{Approximated reliability.}
%	\label{approximatedreliability}
%\end{figure}

%Moreover, even if low bandwidth is used for transmitting safety messages, the approximated reliability is higher than the one of using high bandwidth. 
%It can be concluded that when transmitting packets with larger size, we need to more bandwidth so that the platoon system can achieve a promising reliability performance. 
%\end{figure}

Fig. \ref{controlApproximate} shows the approximated reliability performance under different pairs of control parameters $a$ and $b$ when platoon vehicles are transmitting small packets.
In particular, we assume that both $a$ and $b$ are in the range $(2,4)$ \cite{jin2014dynamics}.
Therefore, by solving the optimization problem in (\ref{newOptim1})--(\ref{newOptimCon4}), we can find the sub-optimal pair of control parameters as $a=2$ and $b=2$.
As shown in Fig. \ref{controlApproximate}, the platoon with the optimized control parameters outperforms platoons with other control parameters.  
In particular, compared with the platoon with control parameters $a=4$ and $b=2$, the reliability gain of the platoon system with the optimized control parameters can be as much as $15$\%.
In addition, the platoon with the optimized control parameters has more flexibility on the platoon spacing. 
For example, to achieve a reliability of $0.9$, the spacing for the platoon with optimized parameters can be at most $25$~m, whereas the spacing for the platoon with $a=3$ and $b=3$ cannot exceed $14$~m.
With more flexibility, the system with the optimized control parameters can tolerate a higher disturbance introduced by rapidly changed road conditions or possible malfunctions of the control system related to the spacing between two consecutive platoon vehicles.

%\begin{figure}[!t]
%	\centering
%	\includegraphics[width=3.5in,height=2.8in]{platoonSize.pdf}
%	\DeclareGraphicsExtensions.
%	\caption{Reliability for platoons with different number of followers.}
%	\label{platoonsize}
%\end{figure}

 \begin{figure}
	\centering	
	\begin{minipage}{0.47\textwidth}
		\centering
		\includegraphics[width=1\linewidth]{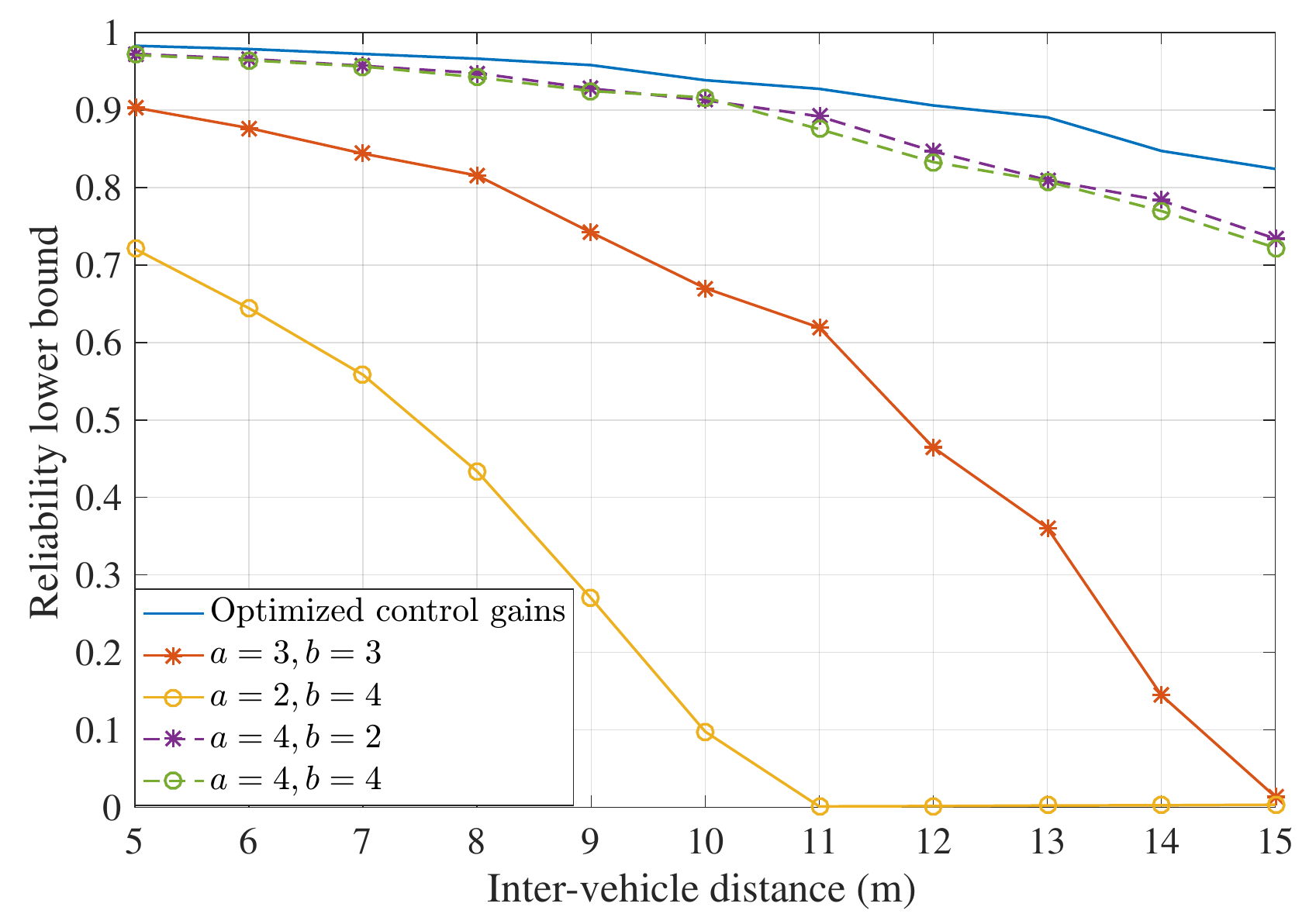}
		\vspace{-0.45in}
		\caption{Optimization design for the control system \\ in Theorem \ref{theorem6}.}
		\label{controlLowerBound}
	\end{minipage}%%%	
	\begin{minipage}{0.47\textwidth}
		\centering
		\includegraphics[width=1\linewidth]{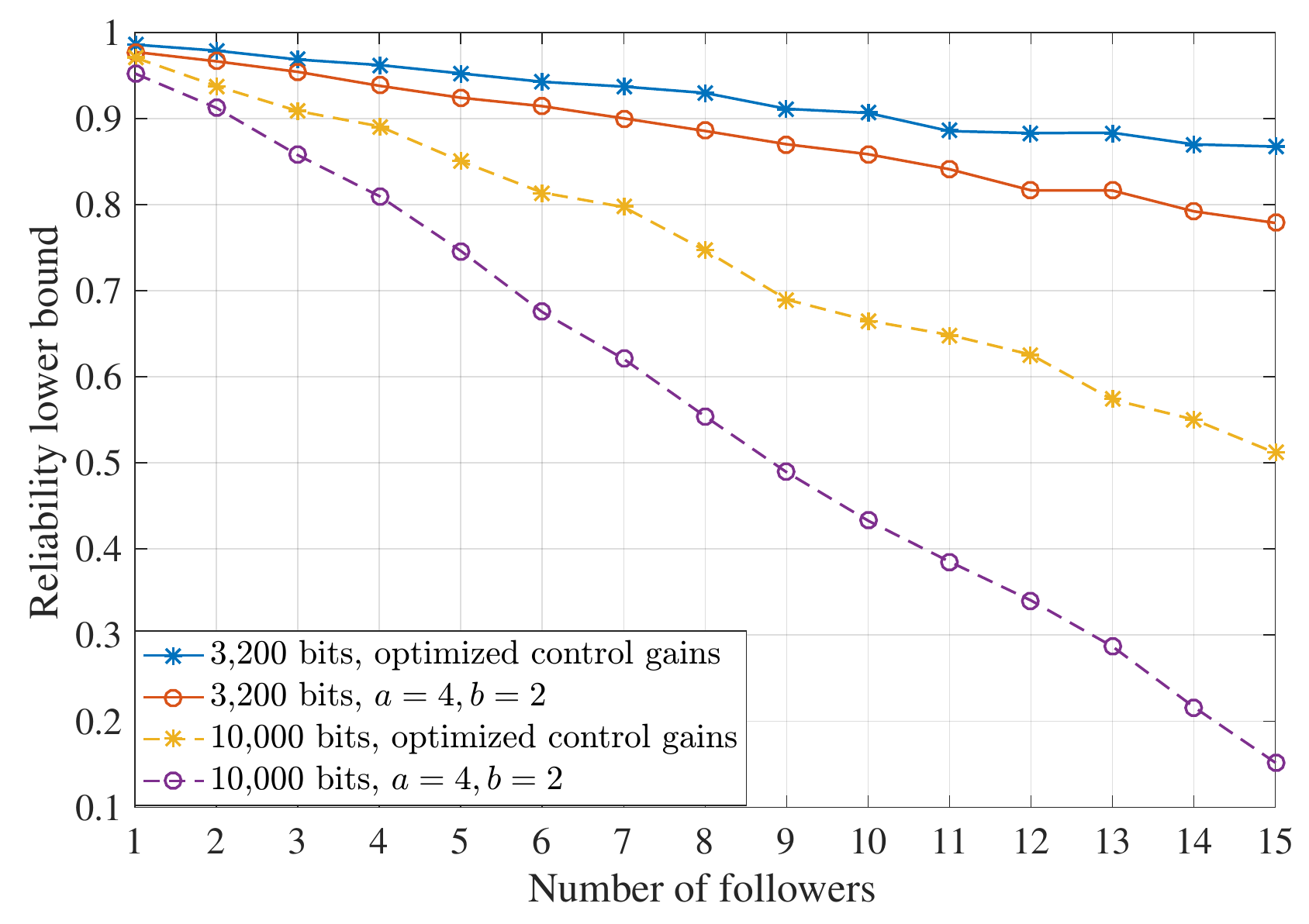}
		\vspace{-0.45in}
		\caption{Reliability for platoons with different number of followers.}
		\label{platoonsize}
	\end{minipage}%%%
	\vspace{-0.45in}
\end{figure}

%\begin{figure}[!t]
%	\centering
%	\includegraphics[width=3.5in,height=2.8in]{approximateOptimize.pdf}
%	\DeclareGraphicsExtensions.
%	\caption{Control design for Theorem \ref{SolveOptimizationProblem}.}
%	\label{controlApproximate}

Fig. \ref{controlLowerBound} shows the reliability lower bounds under different control parameters when platoon vehicles are transmitting small packets.
By using the same parameters as in Fig. \ref{controlApproximate}, both $a$ and $b$ are in the range $(2,4)$.
Based on Corollary \ref{theorem6}, the optimized parameters are $a=2$ and $b=2$, and the performance with optimized parameters is verified in Fig. \ref{controlLowerBound}.  
In particular, the performance gain of choosing the optimized control parameters can be as much as $15$\%, compared with the platoon with control parameters $a=4$ and $b=2$.
To achieve the same reliability, the maximum spacing chosen in Fig. \ref{controlApproximate} has to be much smaller than its counterpart selected in Fig. \ref{controlLowerBound}.
%To achieve the same value of reliability performance, the maximum spacing in Fig. \ref{controlApproximate} for considering the reliability bound would be much smaller than its counterpart in Fig. \ref{controlLowerBound}, when considering approximated reliability.}	
%Comparing Fig.  with Fig. , 
%to achieve the same value of reliability performance, the maximum spacing when considering the reliability bound would be much smaller than the counterpart when considering approximated reliability.}
For example, when we consider the approximated reliability, the inter-vehicle spacing can be as large as $25$~m to realize a reliability $0.9$.
However, when we consider reliability lower bounds, the spacing must be smaller than $12$~m, which is half of the spacing selected when considering the approximated reliability. 
This is due to the fact that when calculating the approximated reliability, we ignore the queuing and processing delays at the processor and the queuing delay at the transceiver, leading to a much larger spacing threshold.

Fig. \ref{platoonsize} shows reliability lower bounds for platoons with different numbers of followers and control parameters. 
We can observe that, as the number of followers increases, 
the reliability of the system (Theorem \ref{theorem4}) decreases.
%reliability performance derived in will decrease. 
This stems from the fact that increasing the number of followers reduces the amount of bandwidth assigned for each V2V link in the platoon.
%when the number of followers increases, the bandwidth assigned to each V2V link in the platoon will decrease. 
%Due to the fixed bandwidth assigned to the platoon, the bandwidth owed by each V2V link in the platoon will decrease when the number of followers is increasing. 
As a result, the transmission rate will decrease, and the performance of the wireless network will degrade. 
Furthermore, according to Fig. \ref{platoonsize}, we can obtain the design guidelines on how to optimize the number of followers in each platoon to realize a target reliability. 
%when designing a platoon system, we need to choose proper number of followers so that the platoon can realize a target reliability performance. 
For example, when transmitting packets with size $3,200$~bits and control gains are $a=4$ and $b=2$, the number of followers should be smaller than $7$ so that the reliability lower bound can be no less than $0.9$. 
In addition, from Fig. \ref{platoonsize}, for different types of packets, we need to choose a proper bandwidth $w$ so as to achieve a satisfactory reliability performance. 
In this regard, by optimizing the design of the control system, we can increase the number of following vehicles and relax the need for a large bandwidth. 
In particular, when transmitting small packets, to realize a $0.9$ reliability performance, the number of followers in the platoon with optimized control parameters can be at most $10$, which is more than the one chosen by the platoon with no optimizations on the control system. By allowing more following vehicles in the platoon, the road capacity can further increase, and, thus, improving the traffic situation.
%we need to determine the proper number of followers so that the platoon can realize can realize the target reliability performance. 
%In particular, to realize that the approximated reliability exceeds $90$\%, the number of followers should be smaller than $8$ when transmitting short safety message.  
\vspace{-0.2in}
\section{Conclusions}
\vspace{-0.05in}
In this paper, we have proposed an integrated communication and control  framework for analyzing the performance and reliability of wireless connected vehicular platoons. 
In particular, we have analyzed plant stability and string stability to derive the maximum wireless system delay that a stable platoon control system can tolerate. 
In addition, we have derived the end-to-end delay, including queuing, processing, and transmission delay, that a packet will encounter in the wireless communication network by using stochastic geometry and queuing theory.
Furthermore, we have conducted theoretical analysis for the reliability of the wireless vehicular platoon, defined as the 
probability of the wireless network meeting the control system’s delay requirements, and derived its lower bounds and approximated expression.
Then, we have proposed two optimization mechanisms to select the control parameters for improving the reliability performance of the wireless network in vehicular platoon systems. 
Simulation results have corroborated the analytical derivations and shown the impact of parameters, such as the density of interfering vehicles, the packet size, and the platoon size, on the reliability performance of the vehicular platoon.
More importantly, the simulation results have shed light on the benefits of the joint control system and wireless network design while providing guidelines to design the platoon system.
In particular, the results provide key insights on how to choose the number of followers, the spacing between two consecutive vehicles, and the control parameters for the control system so as to maintain a stable operation for the autonomous platoon.
Future works will extend the current framework to a more dynamic model with multiple platoons.
%and a trade-off relationship between the control system and vehicular network for platoons.

\vspace{-0.1in}
\appendix
\vspace{-0.0in}
\subsection{Proof of Theorem \ref{theorem1}}
\vspace{-0.1in}
\label{prooffortheorem1}
Since vehicles in the platoon are identical and the channel gains of different V2V links follow the same distribution, plant stability can be guaranteed as long as the delay of each V2V link does not exceed a threshold $\tau_{\text{max}}$. That is, $\tau_{i-1,i}(t)\leq \tau_{\text{max}}, i\in \mathcal{M},$ is the requirement to guarantee the plant stability of the platoon system. 
Thus, we use rewrite (\ref{error}) as follows:\vspace{-0.1in}  
\begin{align}
\dot{\boldsymbol{e}}(t) \stackrel{(a)}{=}& \boldsymbol{M}_{1} \boldsymbol{e} + \sum_{i=1}^{M} \boldsymbol{M}_{2}^{i} \left[\boldsymbol{e}-\int_{-\tau_{\max}}^{0} \dot{\boldsymbol{e}}(t+s)ds \right] \nonumber \\
\stackrel{(b)}{=}& \left(\boldsymbol{M}_{1}+\sum_{i=1}^{M} \boldsymbol{M}_{2}^{i}\right)\boldsymbol{e}-\sum_{i=1}^{M} \boldsymbol{M}_{2}^{i}\int_{-\tau_{\max}}^{0}\boldsymbol{M}_{1}\boldsymbol{e}(t+s)ds \nonumber \\ 
&-\sum_{i=1}^{M} \boldsymbol{M}_{2}^{i}\int_{-\tau_{\max}}^{0}\boldsymbol{M}_{1} \sum_{i'=1}^{M} \boldsymbol{M}_{2}^{i'}e(t+s - \tau_{m'-1,m}(t+s))ds \nonumber \\ 
\stackrel{(c)}{=}& \left(\boldsymbol{M}_{1}+\sum_{i=1}^{M} \boldsymbol{M}_{2}^{i}\right)\boldsymbol{e}-\sum_{i=1}^{M} \boldsymbol{M}_{2}^{i}\int_{-\tau_{\max}}^{0}\boldsymbol{M}_{1}\boldsymbol{e}(t+s)ds \nonumber \\ 
&-\sum_{i=2}^{M}\int_{-\tau_{\max}}^{0}\boldsymbol{M}_{2}^{i}\boldsymbol{M}_{2}^{i-1}e(t+s - \tau_{m'-1,m}(t+s))ds,
\end{align}
where (a) follows the Leibniz--Newton formula, (b) follows from (\ref{error}), and (c) follows from the the fact that $\boldsymbol{M}^{i}_{2}\boldsymbol{M}^{j}_{2}=0$ when $j \neq i-1$. 
To find the value of the threshold $\tau_{\max}$, we use the following candidate Lyapunov function \cite{liu2012consensus}:
$V(\boldsymbol{e}) = \boldsymbol{e}^T \boldsymbol{P} \boldsymbol{e}$, where $\boldsymbol{P}=\boldsymbol{I}_{2M \times 2M}$. 
%Note that, as we can verify that $\boldsymbol{M}_{1}+\sum_{i=1}^{M} \boldsymbol{M}_{2}^{i}$ is a Hurwitz matrix, we can choose $\boldsymbol{P}$ to meet $\boldsymbol{P}(\boldsymbol{M}_{1}+\sum_{i=1}^{M} \boldsymbol{M}_{2}^{i})+(\boldsymbol{M}_{1}+\sum_{i=1}^{M} \boldsymbol{M}_{2}^{i})^T\boldsymbol{P}=-\boldsymbol{I}_{2M \times 2M}$.
According to Lyapunov-Razumikhin theorem introduced in \cite{gu2003stability}, there exists a continuous nondecreasing function $\psi(x)$ that guarantees $\psi(V(\boldsymbol{e})) \geq  V(\boldsymbol{e}(t+t'))$, $t'\in (-\infty,0)$. Then, the time derivative for $V(\boldsymbol{e})$ will be:\vspace{-0.1in}
\begin{align}
\label{Lyapunovderivative}
\dot{V}(\boldsymbol{e}) &=\boldsymbol{e}^T\left(2\left(\boldsymbol{M}_{1}+\sum_{i=1}^{M} \boldsymbol{M}_{2}^{i}\right) \right)\boldsymbol{e} - 2 \boldsymbol{e}^T \sum_{i=1}^{M} \int_{- \tau_{\text{max}} }^{0} \boldsymbol{M}^{i}_{2}\boldsymbol{M}_{1} \boldsymbol{e}(t+s)ds  \nonumber\\
&-2 \boldsymbol{e}^T  \sum_{i=2}^{M}\int_{-\tau_{\max}}^{0}\boldsymbol{M}_{2}^{i}\boldsymbol{M}_{2}^{i-1}e(t+s - \tau_{m'-1,m}(t+s))ds.
\end{align}
Note that for a positive definite matrix $\boldsymbol{\phi}$, we have $2\boldsymbol{v}_{1}^T\boldsymbol{v}_{2}\leq \boldsymbol{v}_{1}^T\boldsymbol{\phi} \boldsymbol{v}_{1} + \boldsymbol{v}_{2}^T\boldsymbol{\phi}^{-1} \boldsymbol{v}_{2}$.		
Thus, let $\boldsymbol{v}_{1}^{T}=-\boldsymbol{e}^T\boldsymbol{M}_{2}^{i}\boldsymbol{M}_{1}$, $\boldsymbol{\phi}=\boldsymbol{I}_{2M \times 2M}$, and $\boldsymbol{v}_{2}=\boldsymbol{e}(t+s)$. Then, the inequality for the second term of the right-hand side in (\ref{Lyapunovderivative}) will be\vspace{-0.05in}
\begin{align}
\label{inequaility1}
-2 \boldsymbol{e}^T \!\sum_{i=1}^{M} \int_{- \tau_{\text{max}} }^{0}\!\!\!\!\! \boldsymbol{M}^{i}_{2}\boldsymbol{M}_{1} \boldsymbol{e}(t\!+\!s)ds\! \leq\! \sum_{i=1}^{M} \left(\int_{- \tau_{\text{max}} }^{0}\!\!\!\!\! \boldsymbol{e}(t\!+\!s)^T\boldsymbol{e}(t\!+\!s)ds \! +\!
\tau_{\text{max}} \boldsymbol{e}^T \boldsymbol{M}_{2}^{i}\boldsymbol{M}_{1} \boldsymbol{M}_{1}^T(\boldsymbol{M}^{i}_{2})^T\boldsymbol{e}\right).
\end{align}
When $V(\boldsymbol{e}(t+s))\leq \psi(V(\boldsymbol{e})) = kV(\boldsymbol{e})$ with $k>1, s\in (- \tau_{\text{max}}, 0)$, (\ref{inequaility1}) can be further simplified as:
$
\label{inequaility2}
-2 \boldsymbol{e}^T  \int_{- \tau_{\text{max}} }^{0} \boldsymbol{M}^{i}_{2}\boldsymbol{M}_{1} \boldsymbol{e}(t+x)dx \leq  \tau_{\text{max}}\sum_{i=1}^{M} \boldsymbol{e}^T (\boldsymbol{M}^{i}_{2}\boldsymbol{M}_{1} \boldsymbol{M}_{1}^T(\boldsymbol{M}^{i}_{2})^T+k \boldsymbol{I}_{2M \times 2M})\boldsymbol{e}.
$
Similarly, we can perform the same steps for the third term on the right-hand side in (\ref{Lyapunovderivative}). 
Finally, we can obtain
$
\dot{V}(\boldsymbol{e}) \leq \boldsymbol{e}^T[2(\boldsymbol{M}_{1}\!+\!\sum_{i=1}^{M} \boldsymbol{M}_{2}^{i})\!+\! \sum_{i=1}^{M}(\tau_{\text{max}}\boldsymbol{M}^{i}_{2}\boldsymbol{M}_{1}\boldsymbol{M}_{1}^T(\boldsymbol{M}^{i}_{2})^T)   +\sum_{i=2}^{M}(\tau_{\text{max}}\boldsymbol{M}^{i}_{2}\boldsymbol{M}^{i-1}_{2}\\(\boldsymbol{M}^{i-1}_{2})^T(\boldsymbol{M}^{i}_{2})^T)+2M\tau_{\text{max}} k\boldsymbol{I}_{2M\times 2M} ]\boldsymbol{e}.
$
Based on the Lyapunov-Razumikhin theorem in \cite{gu2003stability}, if $\dot{V}(\boldsymbol{e}) \!\!\leq \!\!0$, i.e., $\tau_{\text{max}}\!\! \leq\!\! \lambda_{\min}(-2(\boldsymbol{M}_{1}\!+\!\sum_{i=1}^{M} \boldsymbol{M}_{2}^{i}))/\lambda_{\max}\big(\sum_{i=1}^{M}(\boldsymbol{M}_{2}^{i}\boldsymbol{M}_{1}\boldsymbol{M}_{1}^T(\boldsymbol{M}^{i}_{2})^T) +\sum_{i=2}^{M}(\boldsymbol{M}^{i}_{2}\\\boldsymbol{M}^{i-1}_{2}(\boldsymbol{M}^{i-1}_{2})^T (\boldsymbol{M}^{i}_{2})^T)\!+\!2Mk\boldsymbol{I}_{2M\times 2M})\big)$, the system in (\ref{controlLaw}) is asymptotically stable and the augmented error state vector will converge to a zero vector.
Note that, to ensure that $\lambda_{\min}(-2(\boldsymbol{M}_{1}\!+\!\sum_{i=1}^{M} \boldsymbol{M}_{2}^{i}))$ is a real number, the selection of $a_{i}$ and $b_{i}$ should meet $a_{i}^2+b_{i}^2+2a_{i}b_{i}-4a_{i}\geq0$. 
Therefore, to guarantee plant stability, the V2V link delay should not exceed $\tau_{1}=\lambda_{\min}(-2(\boldsymbol{M}_{1}\!+\!\sum_{i=1}^{M} \boldsymbol{M}_{2}^{i}))/\lambda_{\max}\big(\!\sum_{i=1}^{M}(\!\boldsymbol{M}^{i}_{2}\boldsymbol{M}_{1}\boldsymbol{M}_{1}^T(\boldsymbol{M}^{i}_{2})^T)  \!+\!\sum_{i=2}^{M}(\boldsymbol{M}^{i}_{2}\boldsymbol{M}^{i-1}_{2}(\boldsymbol{M}^{i-1}_{2})^T(\boldsymbol{M}^{i}_{2})^T)\!+\!2M k\boldsymbol{I}_{2M\times 2M} \big)$. 
\vspace{-0.35in}
\subsection{Proof of Proposition \ref{theorem2} }
\vspace{-0.1in}
\label{prooffortheorem2}
%To ensure string stability, the magnitude of the transfer function must satisfy $|T(jf)|\leq1$, for $f\in \mathbb{R}^{+}$, where $f$ represents the frequency of sinusoidal excitation signals generated by the leader \cite{swaroop1996string}. 
The magnitude inequality $|T_{i}(jf)|\leq1$ is equivalent to\vspace{-0.1in}
\begin{equation}
\label{magnitude1}
\varGamma_{i}(f) = E_{i} f^{4} + F_{i} f^{2} + G_{i} \geq 0,\vspace{-0.2cm}
\end{equation}
where $E_{i}=0.25(\tau_{i-1,i}(t))^{2}>0$, $F_{i}=(0.5A_i-0.25B_i^2+0.25C_i^2)(\tau_{i-1,i}(t))^{2}+1$, and $G_{i}=C_i^{2}-2A_i-B_i^{2}-2A_iC_i(\tau_{i-1,i}(t))$.
%To solve (\ref{magnitude1}), we need $\varGamma_{i}(f')>0$, where $\frac{d\varGamma_{i}(f)}{df}|_{f=f'}=0$. 
As both $E_{i}, F_{i} > 0, i\in \mathcal{M}$, we can easily find that the delay should satisfy $\tau_{i-1,i}(t) \leq \tau_{2} = \frac{C_i^{2}-2A_i-B_i^{2}}{2A_iC_i}$ so that the string stability of the platoon can be assured.
Moreover, as the associated gains are the same for each platoon vehicle, we can obtain the results in Proposition \ref{theorem2}.
\vspace{-0.3in}
\subsection{Proof of Lemma \ref{lemmaInterference1}}
\vspace{-0.35in}
\label{lemmaInterference1proof}
\begin{align}
\label{SINRCCDF}
\mathcal{L}_{i}^{\text{non-platoon}}(s)=& \mathbb{E}_{\Phi}\left[\exp\left(-s\sum_{j_{1}=1}^{n-1} \sum_{c \in \Phi_{j_{1}}}P_{t} g_{c,i}(t)(d_{c,i}(t))^{-\alpha}\!-\!s
\sum_{j_{2}=n+1}^{N} \sum_{c \in \Phi_{j_{2}}}
P_{t} g_{c,i}(t)(d_{c,i}(t))^{-\alpha}\right)\right] \nonumber \\
=& \mathbb{E}_{\Phi}\left[\exp\left(-s\sum_{j_{1}=1}^{n-1} \sum_{c \in \Phi_{j_{1}}}P_{t} g_{c,i}(t)(d_{c,i}(t))^{-\alpha}\right)\right] \nonumber \\ 
&\times \mathbb{E}_{\Phi}\left[\exp\left(-s\sum_{j_{2}=n+1}^{N} \sum_{c \in \Phi_{j_{2}}}
P_{t} g_{c,i}(t)(d_{c,i}(t))^{-\alpha}\right)\right] \nonumber \\
=& \prod_{j_{1}=1}^{n-1}\mathbb{E}_{\Phi_{j_{1}}}\left[\prod_{c \in \Phi_{j_{1}}}\mathbb{E}_{g_{c,i}}\left( \exp\left(-P_{t} g_{c,i}(t)(d_{c,i}(t))^{-\alpha}\right)\right) \right] \nonumber \\ 
&\times \prod_{j_{2}=n+1}^{N}\mathbb{E}_{\Phi_{j_{2}}}\left[\prod_{c \in \Phi_{j_{2}}}\mathbb{E}_{g_{c,i}}\left( \exp\left(-P_{t} g_{c,i}(t)(d_{c,i}(t))^{-\alpha}\right)\right) \right] \nonumber \\
\stackrel{(a)}{=}& \prod_{j_{1}=1}^{n-1}\mathbb{E}_{\phi_{j_{1}}}\left[\prod_{c \in \phi_{j_{1}}} \frac{1}{1+sP_{t}d_{c,i}^{-\alpha}} \right] \times 
\prod_{j_{2}=n+1}^{N}\mathbb{E}_{\phi_{j_{2}}}\left[\prod_{c \in \phi_{j_{2}}} \frac{1}{1+sP_{t}d_{c,i}^{-\alpha}} \right]
\nonumber \\
\stackrel{(b)}{=}&\prod_{j_{1}=1}^{n-1}\exp\left[-\lambda_{j_{1}}\int_{(n-j_{1})l}^{\infty}\left(1-\frac{1}{1+sP_{t}r^{-\alpha}}\right)\frac{2r}{\sqrt{r^{2}-(n-j_{1})^{2}l^{2}}}dr\right] \nonumber \\
&\times\prod_{j_{2}=n+1}^{N}\exp\left[-\lambda_{j_{2}}\int_{(j_{2}-n)l}^{\infty}\left(1-\frac{1}{1+sP_{t}r^{-\alpha}}\right)\frac{2r}{\sqrt{r^{2}-(j_{2}-n)^{2}l^{2}}}dr\right],
\end{align}
where (a) follows from the assumption of Rayleigh channel where the channel gain follows the exponential distribution.
Also, in (b), $d_{c,i}$ is replaced with $r$.
We have also used the probability generating functional (PGFL) of a Poisson point process \cite{haenggi2012stochastic} and calculated the derivative of $r$ using the relationship between the horizontal distance and vertical distance.
%and the probability generating functional (PGFL) of a PPP is utilized \cite{haenggi2012stochastic} where the derivative of $r$ is calculated using the relationship between the horizontal distance and vertical distance.}

\vspace{-0.3in}
\subsection{Proof of Theorem \ref{meanandvariance}}
\vspace{-0.1in}
\label{SuccessfullyTransmissionProof}
Since the fading channel between vehicles $i-1$ and $i$ is a Nakagami channel, $g_{i,j}$ is a normalized Gamma random variable with parameter $m$.
Based on the distribution of $g_{i,j}$, the CCDF of SINR can be expressed as\vspace{-0.1in}
\begin{align}
\mathbb{F}(\theta)=&\mathbb{P}(\gamma_{i,j} > \theta)= \mathbb{P}\left(\frac{P_{t}g_{i,j}d_{i-1,i}^{-\alpha}}{\sigma^{2}+I_{i}^{\text{non-platoon}}(t)+I_{i}^{\text{platoon}}(t)}> \theta\right) \nonumber \\
=&\mathbb{P}\left(g_{i,j} > \frac{\theta(\sigma^{2}+I_{i}^{\text{non-platoon}}(t)+I_{i}^{\text{platoon}}(t))}{P_{t}}d_{i-1,i}^{-\alpha} \right) \nonumber \\
\stackrel{(a)}{\approx}& 1-\mathbb{E}_{\Phi \cup \Psi}\left[\left(1-\exp\left({\frac{-\eta \theta d_{i-1,i}^{\alpha}(\sigma^{2}+I_{i}^{\text{non-platoon}}(t)+I_{i}^{\text{platoon}}(t))}{P_{t}}}\right)\right)^{m}\right] \nonumber \\
\stackrel{(b)}{=}&\sum_{k=1}^{\beta}(-1)^{k+1}{{\beta}\choose{k}}\mathbb{E}_{\Phi \cup \Psi}\left(\exp\left({\frac{-k\eta \theta d_{i-1,i}^{\alpha}(\sigma^{2}+I_{i}^{\text{non-platoon}}(t)+I_{i}^{\text{platoon}}(t))}{P_{t}}}\right) \right) \nonumber \\
=&\sum_{k=1}^{\beta}(-1)^{k+1}{{\beta}\choose{k}}\exp\left(\frac{-k\eta \theta d_{i-1,i}^{\alpha}}{P_{t}}\sigma^{2}\right)\mathbb{E}_{\Phi}\left(\exp\left(\frac{-k\eta \theta d_{i-1,i}^{\alpha}}{P_{t}}I_{i}^{\text{non-platoon}}(t)\right) \right)  \nonumber \\
&\mathbb{E}_{\Psi}\left(\exp\left(\frac{-k\eta \theta d_{i-1,i}^{\alpha}}{P_{t}}I_{i}^{\text{platoon}}(t)\right) \right) \nonumber \\
\stackrel{(c)}{=}& \sum_{k=1}^{\beta}(-1)^{k+1}{{\beta}\choose{k}}\exp\left(\frac{-k\eta \theta d_{i-1,i}^{\alpha}}{P_{t}}\sigma^{2}\right)\mathcal{L}_{i}^{\text{non-platoon}}\left(\frac{k\eta \theta d_{i-1,i}^{\alpha}}{P_{t}}\right)\mathcal{L}_{i}^{\text{platoon}}\left(\frac{k\eta  \theta d_{i-1,i}^{\alpha}}{P_{t}}\right),
\end{align}
where $\eta=\beta(\beta!)^{-\frac{1}{\beta}}$, (a) is based on the approximation of tail probability of a Gamma function \cite{bai2015coverage}, (b) follows Binomial theorem and the assumption that $m$ is an integer, and the changes in (c) follow the definition of Laplace transform.
Also, we can calculate the PDF of the SINR at vehicle $i$ as
$f(\theta) = \frac{d(1-\mathbb{F}(\theta))}{d\theta} = -\frac{\mathbb{F}(\theta)}{d\theta}$.
Therefore, according to the relationship between the data rate and SINR, we can obtain the mean and variance of service time in (\ref{meanofservicetime}) and (\ref{varianceofservicetime}).
\vspace{-0.3in}
\subsection{Proof of Theorem \ref{theorem4}}
\vspace{-0.1in}
\label{proofforTheorem4}
The first element in the maximization function is actually a lower bound for the reliability, which is proven by \vspace{-0.1in}
\begin{align}
\mathbb{P}(T_{1}+T_{2}\leq \min(\tau_{1},{\tau_{2}})) &= 1- \mathbb{P}(T_{1}+T_{2}\geq \min(\tau_{1},{\tau_{2}})) \nonumber \\
&\stackrel{(a)}{\geq} 1- \frac{\mathbb{E}(T_{1}+T_{2})}{\min(\tau_{1},\tau_{2})} 
= 1- \frac{\bar{T}_{1}+\bar{T}_{2}}{\min(\tau_{1},\tau_{2})},	
\end{align}
where (a) is based on Markov's inequality %$\mathbb{P}(X\geq a)\leq \frac{\mathbb{E}(X)}{a}$ 
\cite{stein2009real}.  		
For the second element in the maximization function, we leverage the Chernoff bound  \cite{hoeffding1963probability} in (a) to obtain another lower bound. Between these two lower bounds, we can always choose the tighter bound to be closer to the reliability of the wireless network, as shown in (\ref{LoweBound1}). 
\vspace{-0.2in}
\vspace{-0.1in}
\subsection{Proof of Corollary \ref{theorem6}}
\vspace{-0.1in}
\label{prooffortheorem6}
To prove Corollary \ref{theorem6}, we first let $\tau_{3}=\frac{\bar{T}_{1}+\bar{T}_{2}}{\min(\tau_{1},{\tau_{2}})}, 0\leq\tau_{3}\leq 1$, and the two functions in the (\ref{optimization2}) can be simplified as \vspace{-0.1in}
\begin{align}
&1- \frac{\bar{T}_{1}+\bar{T}_{2}}{\min(\tau_{1},{\tau_{2}})} = 1-\tau_{3}, \label{theorem61}\\
&1\!\!-\!\!\exp\left(\bar{T}_{1}\!+\!\bar{T}_{2}\!-\!\min(\tau_{1},{\tau_{2}}) \ln\left(\frac{\min(\tau_{1},{\tau_{2}})}{\bar{T}_{1}+\bar{T}_{2}} \right)  \right)=1\!\!-\!\!\exp\left[(\bar{T}_{1}\!+\!\bar{T}_{2})\left(1\!-\!\frac{1}{\tau_{3}}\ln(\frac{1}{\tau_{3}})\right)\right] \label{theorem62},
\end{align}
where both (\ref{theorem61}) and (\ref{theorem62}) are decreasing functions in terms of $\tau_{3}$. 
Note that finding the maximum value between functions (\ref{theorem61}) and (\ref{theorem62}) is equivalent to maximizing the maximal value of (\ref{theorem61}) and (\ref{theorem62}). 
Therefore, the solution is to minimize $\tau_{3}$, which equals to maximizing $\min(\tau_{1},\tau_{2})$.
In other words, the solutions for the optimization problem in (\ref{newOptim1})--(\ref{newOptimCon4}) apply to the problem for the reliability lower bounds as long as the solutions can guarantee $\bar{T}_{1}+\bar{T}_{2} \leq \min(\tau_{1},\tau_{2})$.

\def\baselinestretch{1.00}
\bibliographystyle{IEEEtran}

%\vspace{-0.4cm}
\end{document}